\documentstyle[11pt]{article}
\newcounter{subequation}[equation]

\makeatletter 

\def\citen#1{%
\edef\@tempa{\@ignspaftercomma,#1, \@end, }
\edef\@tempa{\expandafter\@ignendcommas\@tempa\@end}%
\if@filesw \immediate \write \@auxout {\string \citation {\@tempa}}\fi
\@tempcntb\m@ne \let\@h@ld\relax \def\@citea{}%
\@for \@citeb:=\@tempa\do {\@cmpresscites}%
\@h@ld}
\def\@ignspaftercomma#1, {\ifx\@end#1\@empty\else
   #1,\expandafter\@ignspaftercomma\fi}
\def\@ignendcommas,#1,\@end{#1}
\def\@cmpresscites{%
 \expandafter\let \expandafter\@B@citeB \csname b@\@citeb \endcsname
 \ifx\@B@citeB\relax 
    \@h@ld\@citea\@tempcntb\m@ne{\bf ?}%
    \@warning {Citation `\@citeb ' on page \thepage \space undefined}%
 \else
    \@tempcnta\@tempcntb \advance\@tempcnta\@ne
    \setbox\z@\hbox\bgroup 
    \ifnum0<0\@B@citeB \relax
       \egroup \@tempcntb\@B@citeB \relax
       \else \egroup \@tempcntb\m@ne \fi
    \ifnum\@tempcnta=\@tempcntb 
       \ifx\@h@ld\relax 
          \edef \@h@ld{\@citea\@B@citeB }%
       \else 
          \edef\@h@ld{\hbox{--}\penalty\@highpenalty
            \@B@citeB }%
       \fi
    \else   
       \@h@ld\@citea\@B@citeB
       \let\@h@ld\relax
 \fi\fi%
 \def\@citea{,\penalty\@highpenalty\hskip.13em plus.1em minus.1em}%
}
\def\@citex[#1]#2{\@cite{\citen{#2}}{#1}}%
\def\@cite#1#2{\leavevmode\unskip
  \ifnum\lastpenalty=\z@\penalty\@highpenalty\fi
  \ [{\multiply\@highpenalty 3 #1
      \if@tempswa,\penalty\@highpenalty\ #2\fi 
    }]\spacefactor\@m}
\def\thesubequation{\theequation\@alph\c@subequation}
\def\@subeqnnum{{\rm (\thesubequation)}}
\def\slabel#1{\@bsphack\if@filesw {\let\thepage\relax
   \xdef\@gtempa{\write\@auxout{\string
      \newlabel{#1}{{\thesubequation}{\thepage}}}}}\@gtempa
   \if@nobreak \ifvmode\nobreak\fi\fi\fi\@esphack}
\def\subeqnarray{\stepcounter{equation}
\let\@currentlabel=\theequation\global\c@subequation\@ne
\global\@eqnswtrue
\global\@eqcnt\z@\tabskip\@centering\let\\=\@subeqncr
$$\halign to \displaywidth\bgroup\@eqnsel\hskip\@centering
  $\displaystyle\tabskip\z@{##}$&\global\@eqcnt\@ne
  \hskip 2\arraycolsep \hfil${##}$\hfil
  &\global\@eqcnt\tw@ \hskip 2\arraycolsep
  $\displaystyle\tabskip\z@{##}$\hfil
   \tabskip\@centering&\llap{##}\tabskip\z@\cr}
\def\endsubeqnarray{\@@subeqncr\egroup
                     $$\global\@ignoretrue}
\def\@subeqncr{{\ifnum0=`}\fi\@ifstar{\global\@eqpen\@M
    \@ysubeqncr}{\global\@eqpen\interdisplaylinepenalty \@ysubeqncr}}
\def\@ysubeqncr{\@ifnextchar [{\@xsubeqncr}{\@xsubeqncr[\z@]}}
\def\@xsubeqncr[#1]{\ifnum0=`{\fi}\@@subeqncr
   \noalign{\penalty\@eqpen\vskip\jot\vskip #1\relax}}
\def\@@subeqncr{\let\@tempa\relax
    \ifcase\@eqcnt \def\@tempa{& & &}\or \def\@tempa{& &}
      \else \def\@tempa{&}\fi
     \@tempa \if@eqnsw\@subeqnnum\refstepcounter{subequation}\fi
     \global\@eqnswtrue\global\@eqcnt\z@\cr}
\let\@ssubeqncr=\@subeqncr
\@namedef{subeqnarray*}{\def\@subeqncr{\nonumber\@ssubeqncr}\subeqnarray}
\@namedef{endsubeqnarray*}{\global\advance\c@equation\m@ne%
                           \nonumber\endsubeqnarray}

\@addtoreset{equation}{section}
\renewcommand{\theequation}{\thesection.\arabic{equation}}

\makeatother

\def\dalemb#1#2{{\vbox{\hrule height .#2pt
        \hbox{\vrule width.#2pt height#1pt \kern#1pt
                \vrule width.#2pt}
        \hrule height.#2pt}}}
\def\square{\mathord{\dalemb{6.8}{7}\hbox{\hskip1pt}}}

\let\a=\alpha \let\b=\beta \let\g=\gamma  \let\e=\epsilon
  \let\q=\theta \let\iota=\iota 
  \let\n=\nu   
 \let\t=\tau    
         
 \let\Pi=\Pi \let\Sigma=\Sigma

\def\nn{\nonumber} \def\bd{\begin{document}} \def\ed{\end{document}}
\def\ds{\documentstyle} \let\fr=\frac \let\bl=\bigl \let\br=\bigr
\let\Br=\Bigr \let\Bl=\Bigl 
\let\bm=\bibitem
\let\na=\nabla
\let\pa=\partial \let\ov=\overline
\def\ie{{\it i.e.\ }} 
\newcommand{\be}{\begin{equation}} 
\newcommand{\ee}{\end{equation}}
\def\ba{\begin{array}}
\def\ea{\end{array}}
\def\ft#1#2{{\textstyle{{\scriptstyle #1}\over {\scriptstyle #2}}}}
\def\fft#1#2{{#1 \over #2}}
\def\del{\partial}
\def\st#1{{\scriptstyle #1}}
\def\sst#1{{\scriptscriptstyle #1}}
\def\oneone{\rlap 1\mkern4mu{\rm l}}
\def\e7{E_{7(+7)}}
\def\td{\tilde}
\def\wtd{\widetilde}
\def\im{{\rm i}}
\def\bog{Bogomol'nyi\ }
\def\q{{\tilde q}}
\def\hast{{\hat\ast}}
\def\0{{\sst{(0)}}}
\def\1{{\sst{(1)}}}
\def\2{{\sst{(2)}}}
\def\3{{\sst{(3)}}}
\def\4{{\sst{(4)}}}
\def\5{{\sst{(5)}}}
\def\6{{\sst{(6)}}}
\def\7{{\sst{(7)}}}
\def\8{{\sst{(8)}}}
\def\n{{\sst{(n)}}}
\def\tV{{\wtd V}}
\def\cF{{\cal F}}
\def\cA{{\cal A}}
\def\hA{{\hat{\cal A}}}
\def\td{\tilde}
\def\wtd{\widetilde}
\def\ep{\epsilon}
\def\Z{\rlap{\sf Z}\mkern3mu{\sf Z}}
\def\R{\rlap{\rm I}\mkern3mu{\rm R}}
\def\I{{\sst I}}
\def\J{{\sst J}}
\def\K{{\sst K}}
\def\cM{{\cal M}}
\def\T{{\rm T}}
\def\vp{{\varphi}}
\def\sA{{\sst A}}
\def\sB{{\sst B}}
\def\sC{{\sst C}}
\def\sD{{\sst D}}
\def\wb#1{{{\bar #1}}}

\def\cramp{\medmuskip = 2mu plus 1mu minus 2mu}
\def\cramper{\medmuskip = 2mu plus 1mu minus 2mu}
\def\crampest{\medmuskip = 1mu plus 1mu minus 1mu}
\def\uncramp{\medmuskip = 4mu plus 2mu minus 4mu}

\newcommand{\ho}[1]{$\, ^{#1}$}
\newcommand{\hoch}[1]{$\, ^{#1}$}
\newcommand{\bea}{\begin{eqnarray}} 
\newcommand{\eea}{\end{eqnarray}}
\newcommand{\bsea}{\begin{subeqnarray}}
\newcommand{\esea}{\end{subeqnarray}}
\newcommand{\ra}{\rightarrow}
\newcommand{\lra}{\longrightarrow}
\newcommand{\Lra}{\Leftrightarrow}
\newcommand{\ap}{\alpha^\prime}
\newcommand{\bp}{\tilde \beta^\prime}
\newcommand{\tr}{{\rm tr} }
\newcommand{\Tr}{{\rm Tr} } 
\newcommand{\NP}{Nucl. Phys. }
\newcommand{\tamphys}{\it Center for Theoretical Physics,
Texas A\&M University, College Station, Texas 77843}
\newcommand{\ens}{\it Laboratoire de Physique Th\'eorique de l'\'Ecole
Normale Sup\'erieure\hoch{2,3}\\
24 Rue Lhomond - 75231 Paris CEDEX 05}
\newcommand{\sissa}{\it SISSA, Via Beirut No. 2-4, 34013 Trieste, 
Italy\hoch{2}}

\newcommand{\auth}{H. L\"u\hoch{\$\ddagger}, 
C.N. Pope\hoch{\dagger\S\,1} and K.S. Stelle\hoch{\star\sharp}}

\begin{document}
\begin{flushright}
\hfill{CERN-TH/98-303, \ \ CTP TAMU-36/98, \ \  
Imperial/TP/97-98/77, \\
 LPTENS-98/39,\ \ UPR/0819-T}\\
\hfill{\bf hep-th/9810159}\\
\hfill{October 1998}\\
\end{flushright}


\begin{center}
{\Large {\bf M-theory/heterotic Duality: a Kaluza-Klein Perspective}}

\vspace{15pt}

\auth

\vspace{10pt}
{\hoch{\star} \it The Blackett Laboratory, Imperial College\hoch{2}\\
Prince Consort Road, London SW7 2BZ, UK}

\vspace{7pt}
{\hoch{\sharp} \it TH Division, 
CERN, CH-1211 Geneva 23, Switzerland}

\vspace{7pt}
{\hoch{\dagger}\tamphys}

\vspace{7pt}
{\hoch{\$}\it Department of Physics, University of Pennsylvania,
Philadelphia, PA 19104}

\vspace{7pt}
{\hoch{\ddagger}\ens}

\vspace{7pt}
{\hoch{\S}\sissa}

\vspace{20pt}

\underline{ABSTRACT}
\end{center}

     We study the duality relationship between M-theory and heterotic
string theory at the classical level, emphasising the transformations
between the Kaluza-Klein reductions of these two theories on the K3 and
$T^3$ manifolds. Particular attention is devoted to the corresponding
structures of $\sigma$-model cosets and the correspondence between the
$p$-brane charge lattices.  We also present simple and detailed
derivations of the global symmetries and coset structures of the
toroidally-compactified heterotic theory in all dimensions $D\ge3$,
making use of the formalism of solvable Lie algebras.

{\vfill\leftline{}\vfill
\vskip  10pt
\footnoterule
{\footnotesize 
        \hoch{1}        Research supported in part by DOE 
grant DE-FG03-95ER40917. \vskip -12pt}  \vskip  14pt
{\footnotesize
        \hoch{2} Research supported in part by the EC under TMR
contract ERBFMRX-CT96-0045. \vskip -12pt} \vskip 14pt
{\footnotesize
        \hoch{3} Unit\'e Propre du Centre National de la Recherche
Scientifique, associ\'ee \`a l'\'Ecole Normale Sup\'erieure
\vskip -12pt} \vskip 10pt
{\footnotesize \hoch{\phantom{3}} et \`a l'Universit\'e de Paris-Sud.
\vskip -12pt} \vskip 10pt}

\pagebreak
\setcounter{page}{1}

\tableofcontents
\addtocontents{toc}{\protect\setcounter{tocdepth}{2}}
\newpage

\section{Introduction\label{sec:intro}}

     The duality relations \cite{ht,wit1,hwit1,hwit2} between the
heterotic string theory and M-theory are perhaps the most surprising
of the web of dualities that is now seen to underpin the search for a
satisfactory nonperturbative formulation of a quantum theory of
gravity and everything else. One of the most striking features of this
web of dualities is the degree to which such nonperturbative relations
can be seen already in nonlinear features of the classical or
semiclassical field theory limits of the underlying quantum
theories. In the present paper, we shall investigate in some detail
the classical Kaluza-Klein relations between the heterotic theory and
$D=11$ supergravity, which is the field-theory limit of
M-theory. Specifically, we shall consider the relation between $D=11$
supergravity compactified on a K3 manifold and the heterotic theory
compactified on $T^3$.

     A number of general features of the M-theory/heterotic
correspondence were originally detailed in Refs \cite{ht,wit1}. In the
present paper, we shall study some of the apsects of the Kaluza-Klein
reduction procedure in more detail, to reveal further aspects of the
correspondences between the two theories. We begin in sections
\ref{sec:globsym} and \ref{sec:solvable} by considering the nonlinear
$\sigma$-model structure of the dimensionally-reduced heterotic
string, focusing in particular on the relation between group-theoretic
coset descriptions of these $\sigma$-models and embeddings of the
$\sigma$-models into linearly realised representations of the
numerator symmetry groups, combined with appropriate invariant
constraints. Generalising the Borel-subalgebra constructions of
analogous $\sigma$-models in the maximally supersymmetric
supergravities \cite{cjlp1,trombone}, we find that the appearance of
solvable subalgebras \cite{iwa,alek,fre1,fre2,fre3,fre4} of the
heterotic theory duality symmetries significantly simplifies the
derivation of explicit parametrisations for the corresponding
heterotic $\sigma$-model cosets.  We include a relatively digestible
discussion of solvable Lie algebras, taking examples from some of the
simpler global symmetry groups arising in the heterotic
compactifications to illustrate the essential ideas.

     In section \ref{sec:k3}, we analyse the dimensional reduction of
M-theory on a K3 manifold, using in particular the approximate
description of a K3 manifold as $T^4/Z_2$, with the 16 orbifold
singularities of this construction blown up by the cutting and pasting
in of 16 Eguchi-Hanson instantons \cite{page}. This analysis will
allow us to establish a detailed correspondence between the fields of
M-theory and the heterotic theory, and in particular to establish the
relationship between the couplings of dilatonic scalars in the two
theories.

     In section \ref{sec:chargelattice}, we proceed to establish the
correspondence between the lattices of $p$-brane charges in the two
theories. This correspondence reconfirms the structure of the M-theory
charge lattice that had originally been derived using duality
relations \cite{schwarzp,alwis} and the special properties of
``scale-setting'' $p$-brane species \cite{dirac}. The paper concludes
with a discussion of the charge orbits containing $p$-brane solutions
supported purely by the Yang-Mills sector. Given the currently
anticipated general relationship between string theory states and
semiclassical solutions, such solutions are required to correspond to
the short massive multiplets of Yang-Mills sector states associated to
fields in the heterotic Lagrangian acquiring masses from the Higgs
effect for a generic heterotic vacuum. A perhaps unsettling feature of
the only available supersymmetric $p$-branes supported by the
Yang-Mills sector is that they have naked singularities. We shall
explain how these naked singularities appear as an artefact of
dimensional reduction from wave-like solutions using singular Killing
vectors.

     In the appendices, we give some details of the toroidal
dimensional reduction of the heterotic theory and also some additional
embeddings of coset models into constrained linear realisations of
symmetry groups.

\section{Global symmetries of the
heterotic string on $T^n$\label{sec:globsym}}

\subsection{$D=9$ heterotic string\label{ssec:d=9het}}

    Taking the general $T^n$ reduction of the heterotic theory given
in Appendix \ref{app:hetred}, and specialising to the case of
reduction to $D=9$ on a single circle, we find, after rotating the
dilatonic scalars so that
\be
\phi_1 = \ft1{2\sqrt2} \, \varphi - \ft12{\sqrt\ft72}\, \phi\ ,\qquad
\phi_2= \ft12{\sqrt\ft72}\, \varphi + \ft1{2\sqrt2}\, \phi 
\ ,\label{dilatonredef}
\ee
 that the nine-dimensional Lagrangian is given by
\bea
e^{-1}\, {\cal L}_9 &=& R -\ft12 (\del\phi)^2
-\ft12(\del\varphi)^2 - \ft12 e^{\sqrt2\varphi}\, \sum_I 
 (\del B_\0^I)^2 -\ft1{12} e^{-\sqrt{\ft87}\phi}\, (F_\3)^2
\nn\\
&& -\ft14 e^{-\sqrt{\ft27}\phi}\, \Big( e^{\sqrt2 \varphi}\,
(F_\2)^2 + e^{-\sqrt2 \varphi}\, (\cF_\2)^2 + \sum_I
(G_\2^I)^2 \Big)\ ,\label{d9lag}
\eea
 where the $\scriptstyle I$ index labels the 16 unbroken
Cartan-subalgebra gauge-group generators for a generic
``fully-Higgsed'' vacuum configuration. From (\ref{newfields}), and
dropping the primes on the potentials, the field strengths are given
by
\bea
&& F_\3 = dA_\2 + \ft12 B_\1^\I\, dB_\1^\I - \ft12 \cA_\1\, dA_\1
-\ft12 A_\1\, d\cA_\1\ ,\nn\\
&& \cF_\2 = d\cA_\1\ ,
\qquad G_\2^\I = dB_\1^\I\, + B_\0^\I\, \cA_\1\ ,\nn\\
&&F_\2 = dA_\1 + B_\0^\I\, dB_\1^\I + \ft12 B_\0^\I\, B_\0^\I\, 
    d\cA_\1 \ , \qquad 
\eea
 the nine-dimensional string coupling constant is given by $\lambda_9 =
e^{\sqrt{7/8}\, \phi}$.

The symmetry group of the scalar manifold, parametrised by the
dilatons $\phi$ and $\varphi$, and the axions $B_\0^I$, is easily
analysed.  Let us, for greater generality, consider the case where
there are $N$ abelian vectors potentials $B_\1^I$ ($1\le I\le N$) in
$D=10$, rather than the particular case $N=16$ arising in the
heterotic string.  We now introduce the set of $N+2$ fields $X, Y,
Z^I$ in $\R^{1, N +1}$, defined by
\bea
X+Y= e^{\ft1{\sqrt2}\varphi}\ ,\qquad X-Y = e^{-\ft1{\sqrt2}\varphi} +
\ft12 B_\0^I\, B_\0^I\, e^{\ft1{\sqrt2}\varphi} \ ,\qquad Z^I =
\ft1{\sqrt2}\, B_\0^I\, e^{\ft1{\sqrt2}\varphi}\ .
\eea
 It is evident that these satisfy the constraint 
\be
X^2-Y^2-Z^I\, Z^I=1\ ,\label{constraint}
\ee
 and that the scalar part of the Lagrangian (\ref{d9lag}) can be
written as
\be
e^{-1}\, {\cal L}_{\rm scalar} = (\del X)^2 -(\del Y)^2 -(\del Z^I)^2
 -\ft12 (\del\phi)^2 \ ,\label{d9scal}
\ee
subject to the constraint (\ref{constraint}).  Thus we see that the
Lagrangian and the constraint are invariant under global $O(1,N+1)$
transformations, which act by matrix multiplication on the column
vector $(X,Y,Z^I)$, and also that the Lagrangian is invariant under
constant shifts of $\phi$.  Thus the symmetry of the scalar Lagrangian
is $O(1,N+1)\times \R$.

    We find that in terms of the fields $(X, Y, Z^I)$, the Lagrangian
(\ref{d9lag}) can be written as
\bea
e^{-1}\, {\cal L}_9 \!\!\! &=& \!\!\! R + {\cal L}_{\rm scalar} -\ft12
e^{-\sqrt{\ft87}\phi} \, (F_\3)^2 \label{d9lag2}\\
\!\!\! && \!\!\! + \ft14 e^{-\sqrt{\ft27}\phi}\, \Big((dA_X)^2 -
(dA_Y)^2 - (dB_\1^I)^2 
 - 2 (X\, dA_X + Y\, dA_Y + Z^J\, dB_\1^J)^2 \Big)\ ,
\nn
\eea
 where we define
\be
A_X = \ft1{\sqrt2}\, (A_\1 + \cA_\1)\ ,\qquad A_Y = \ft1{\sqrt2}\, 
(A_\1 - \cA_\1)\ .
\ee
In terms of these redefined potentials, the 3-form field $F_\3$
becomes
\be
F_\3 = dA_\2 - \ft12 A_X\, dA_X +\ft12 A_Y\, dA_Y +\ft12 B_\1^I\,
dB_\1^I\ .
\ee
It is now manifest that if the $O(1,N+1)$ transformations act on the
column co-vector $(A_X,A_Y,B_\1^I)$ of 1-form potentials at the same
time as they act on $(X,Y,Z^I)$, then the entire Lagrangian
(\ref{d9lag2}) is invariant.  To be precise, if $\Lambda$ is an
$O(1,N+1)$ group element then the Lagrangian is invariant under the
global transformation
\be
\pmatrix{X\cr Y\cr Z^\I} \longrightarrow \Lambda\,\pmatrix{X\cr Y\cr
  Z^\I}\ ,\qquad
\pmatrix{A_X\cr A_Y\cr B^\I} \longrightarrow (
\Lambda^\T)^{-1}\,\pmatrix{A_X\cr A_Y\cr  B^\I}\ .\label{o2ntrans}
\ee
Note that $F_\3$ is a singlet.  In particular, this completes our
demonstration that the dimensional reduction of the heterotic theory
with $U(1)^{16}$ gauge fields gives a theory with an $O(1,17)$ global
symmetry in $D=9$.  There is in addition the previously-mentioned $\R$
factor also, corresponding to a constant shift of the dilaton $\phi$,
accompanied by appropriate rescalings of the potentials.

\subsection{Geometry of $O(p,q)/(O(p)\times O(q))$ 
cosets\label{ssec:grassman}}

In the previous subsection, we showed how the global symmetry of the
heterotic theory when reduced on $S^1$ could be understood from a
geometrical point of view.  In fact, more generally, we showed that
the reduction of the bosonic sector of ten-dimensional $N=1$
supergravity coupled to $N$ abelian gauge fields gives rise to a
nine-dimensional theory with an $O(1,N+1)\times\R$ global symmetry.
The $O(1,N+1)$ symmetry could be understood geometrically as a set of
linear transformations on an $(N+2)$-vector of coordinates
$(X,Y,Z^\I)$ on $\R^{1,N+1}$ with the Minkowski metric $\eta={\rm
diag}\, (-1,1,1,\ldots,1)$.  The scalar manifold is therefore the
coset space $O(1,N+1)/O(N+1)$ (together with an extra trivial $\R$
factor).

     In this subsection, we shall consider the geometry of the coset
spaces that arise in general toroidal dimensional reductions of the
heterotic theory.  In fact, we shall consider more generally the
geometrical construction of the entire class of coset spaces
\be
\fft{O(p,q)}{O(m,q-n)\times O(p-m,n)}\ ,\label{cosets}
\ee
 where $0\le m\le p$ and $0\le n\le q$.  

     To begin, we introduce the indefinite-signature flat metric
\be
\eta = {\rm diag}\, (\underbrace{-1,-1,\ldots,-1}_p ,
\underbrace{1,1,\ldots, 1}_q)\ .\label{mink}
\ee
 Group elements $W$ in $O(p,q)$ then, by definition, satisfy
\be
W^\T\, \eta\, W = \eta\ .\label{opqel}
\ee
 One can also then impose the additional $O(p,q)$-covariant constraint
\be
W^\T=W\ .\label{sym}
\ee
 Note that matrices $W$ satisfying these two conditions do
not form a group; rather, as we shall now show, they decompose into
orbits described by cosets of the form (\ref{cosets}).\nopagebreak  

    To make this more precise, we shall consider the orbits 
of matrices $W$ satisfying (\ref{opqel}, \ref{sym}),
under the $O(p,q)$ transformations
\be
W\longrightarrow \Lambda^\T\, W\, \Lambda\ .
\ee\goodbreak
Noting that any non-degenerate symmetric matrix $S$ can be
diagonalised under the action $S\rightarrow \Lambda^\T\, S\, \Lambda$
where $\Lambda$ is some $O(p,q)$ matrix, we see that every $O(p,q)$
orbit for $W$ satisfying (\ref{opqel}, \ref{sym}) passes through a
point where $W$ is a diagonal matrix.  For diagonal matrices,
(\ref{opqel}) implies directly that the diagonal elements are all
$\pm1$.  Thus we may characterise the $O(p,q)$ orbits by a fiducial
matrix
\be
W_0 = {\rm diag}\, (\underbrace{-1,-1,\ldots,-1}_m, 
 \underbrace{1,1,\ldots,1}_{p-m}, \underbrace{-1,-1,\ldots,-1}_n, 
\underbrace{1,1,\ldots,1}_{q-n})\ ,\label{fiducial}
\ee
where $0\le m\le p$ and $0\le n\le q$.  The specific distributions of
the $+1$ and $-1$ eigenvalues within the $p\times p$ timelike and
$q\times q$ spacelike subspaces may be modified by further $O(p)\times
O(q)$ transformations.  The fiducial matrix $W_0$ in (\ref{fiducial})
is therefore representative of a class of equivalent matrices.  Note,
however, that no $O(p,q)$ transformation is able to exchange
eigenvalues between the timelike and spacelike sectors, since the
corresponding eigenvectors would also have to be exchanged, and this
is impossible since $O(p,q)$ transformations preserve the norms of
vectors, while timelike and spacelike vectors have norms of opposite
signs.  Thus, the numbers $m$ and $n$ in (\ref{fiducial}) are
$O(p,q)$-invariant.  These numbers will determine the denominator
groups $K$ in the coset spaces $O(p,q)/K$.

    For a given fiducial matrix $W_0$, the denominator group $K$ is
easily determined, since it is nothing but the stability subgroup of
$O(p,q)$ that leaves $W_0$ invariant.  From (\ref{mink}) and
(\ref{fiducial}), we can see that $W_0$ will be left invariant by the
subgroup
\be
K=O(m,q-n)\times O(p-m,n) \label{ksg}
\ee
 of $O(p,q)$.  Thus the orbit for this particular fiducial matrix $W_0$
is the coset space (\ref{cosets}).  Note that all points on the given
orbit have matrices $W$ that satisfy the trace condition
\be
\tr(W\, \eta) = q-p+2(m-n)\ .\label{trace}
\ee
 This follows from the fact that $\tr(W\, \eta)$ is manifestly  
$O(p,q)$-invariant, and that $\tr(W_0\, \eta)$ can be determined by
inspection from (\ref{mink}) and (\ref{fiducial}).  Note that if we
consider the particular fiducial matrix
\be
W_0 = {\rm diag}\, (1,1,\ldots,1)\ ,\label{identity}
\ee
 corresponding to $m=n=0$, then the $O(p,q)$ orbits will describe the
coset
\be
\fft{O(p,q)}{O(p)\times O(q)}\ .\label{opqcoset}
\ee
 It is cosets of this type that will be principally relevant in our
subsequent discussion.  We discuss some further examples of classes of
cosets in Appendix \ref{app:morecos}.

\subsection{Global $O(10-D,10-D+N)$ symmetries 
 from dimensional reduction\label{ssec:globalsym}}

     We saw in subsection \ref{ssec:d=9het} that the bosonic sector of
ten-dimensional simple supergravity coupled to $N$ abelian gauge
multiplets gives rise, when reduced on $S^1$, to a theory with a
global $O(1,N+1)\times \R$ symmetry in nine-dimensions.  In order to
bring out the geometrical structure, we have exploited the fact that
the coset $O(1,N+1)/O(N+1)$ could be viewed as a hypersurface in the
flat Minkowski-signature space $\R^{1,N+1}$. In this subsection, we
shall extend the discussion to the lower-dimensional theories obtained
by compactifying on $T^{10-\sst D}$ instead. In these cases, one
expects the global symmetry group to be $O(10-D,10-D+N)\times \R$ when
$D\ge5$, while in $D=4$ and $D=3$ one expects $O(6,N+6)\times
SL(2,\R)$ and $O(8,N+8)$ respectively.\footnote{These symmetry
enlargements in $D=4$ and $D=3$ result from dualising the 2-form or
1-form potentials to give additional axionic scalars.  If one leaves
them undualised instead, then the $D\ge5$ discussion of global
symmetries extends uniformly to include these dimensions too.} Here,
we shall make use of the more general discussion of coset spaces given
in subsection \ref{ssec:grassman} in order to give a geometrical
interpretation of the symmetries of the $D$-dimensional theories.

   Our starting point is the $D$-dimensional Lagrangian
(\ref{ddimhet}), obtained by dimensional reduction from $D=10$.  We
use the expressions (\ref{newfields}) for the 3-form and 2-form field
strengths, which were obtained after making the field redefinitions
(\ref{dredefs}).  First, we note that the particular linear
combination $\vec a_1\cdot\vec\phi$ of dilatons that couples to the
3-form field strength is decoupled from all the axionic scalars
$A_{\0\a\b}$, $\cA^\a_{\0\b}$ and $B^\I_{\0\a}$.  In other words, the
dot products $\vec a_{1}\cdot\vec a_{1\a\b}$, $\vec a_1\cdot\vec
b_{\a\b}$ and $\vec a_1\cdot\vec c_\a$ all vanish.  It is therefore
natural to perform a rotation on the dilatons so that the combination
$\phi= -\sqrt{(D-2)/8}\, \vec a_1\cdot\vec\phi$ is separated from the
rest.  After this rotation, one expects that the Lagrangian should be
expressible in the form
\be
e^{-1}\, {\cal L}_{\sst D} = R - \ft12(\del\phi)^2 + \ft14
\tr(\del\cM^{-1}\, \del \cM) -\ft1{12}  
e^{-\sqrt{8/(D-2)}\, \phi}\, F_\3^2
-\ft14 e^{-\sqrt{2/(D-2)}\, \phi}\, H_\2^\T\, \cM\, H_\2
\ ,\label{dlagm}
\ee
where $\cM$ is a square matrix of dimension $(20-2D+N)$ that is
parametrised by the rest of the dilatonic scalars and the axionic
scalars, and $H_\2=dC_\1$ is a column vector formed from the exterior
derivatives of the 1-form potentials.  The NS-NS 3-form $F_\3$ is
coupled only to the dilaton $\phi$, implying that the $D$-dimensional
string coupling is $\lambda_{\sst D} = \exp(\sqrt{(D-2)/8}\, \phi)$.

To show that the Lagrangian (\ref{ddimhet}) is invariant under
$O(10-D,10-D+N)$, and that its scalar manifold is the coset space
$O(10-D,10-D+N)/(O(10-D)\times O(10-D+N))$ (together with a further
$\R$ factor for the decoupled scalar $\phi$), we need only show that
it can indeed be written in the form (\ref{dlagm}), and that $\cM$
satisfies the constraints (\ref{opqel}, \ref{sym}) and it is equal to
the identity for some special set of values of the scalar fields.
Furthermore, the number of independent scalar fields in $\cM$ is equal
to the dimension of the coset. The conditions (\ref{opqel}, \ref{sym})
ensure that the symmetry is $O(10-D,10-D+N)$, while the occurrence of
the special point $\cM=\oneone$ ensures that the orbits are those
containing the fiducial point $W_0$ given in (\ref{opqcoset}), leading
to the coset structure (\ref{opqcoset}).

     The easiest way to determine the $\cM$ matrix from
(\ref{ddimhet}) is by studying the kinetic terms for the 2-form field
strengths, and comparing them with the expression $H^\T\, \cM\, H$ in
(\ref{dlagm}).  From (\ref{ddimhet}), we see that, after making the
rotation of dilatons described above, we must have
\be
H_\2^\T\, \cM\, H_\2 = \sum_\a e^{\vec c_\a\cdot\vec\phi} \, (F_{\2\a})^2 
 + \sum_\a e^{-\vec c_\a\cdot\vec\phi} \, (\cF^\a_\2)^2
 + \sum_\I (G_\2^\I)^2\ ,\label{equal}
\ee
since we have $\vec c_\a = \vec a_{1\a} -\ft12 \vec a_1 = -\vec b_\a +
\ft12 \vec a_1$.  It is convenient to express the matrix $\cM$, which
should be symmetric, as $\cM = {\cal V}^\T\, {\cal V}$, so that the 
left-hand side
of (\ref{equal}) can be written as $({\cal V}\, H_\2)^\T\, 
({\cal V}\, H_\2)$.  We
can then read off the matrix ${\cal V}$ by inspecting the expressions in
(\ref{newfields}) for the 2-form field strengths.  It is convenient to
order the various 1-form potentials so that the column vector $C_\1$
is given by
\be
C_\1=
 \pmatrix{A_{\1\a} \cr B^\I_\1 \cr \cA_\1^\a}\ ,
\qquad H_\2 = dC_\1 \ ,\label{Cdef}
\ee
where it is understood that the indices $\a$ and $I$ increase as one
descends downwards through the sets of fields in the column vector.
We then find that ${\cal V}$ is given by
\medskip
\be
{\cal V}= 
\left(
\begin{array}{c|c|c}
e^{\fft12\vec c_\a\cdot\vec\phi} \, \g^\b{}_\a  
& e^{\fft12\vec c_\a\cdot\vec\phi}\, \g^\g{}_\a\, B^\I_{\0 \gamma}   
& e^{\fft12\vec c_\a\cdot\vec\phi}\, \g^\g{}_\a\, (A_{\0\g\b} + \ft12
B^\I_{\0\g}\,  B^\I_{\0\b})   \\ \hline
0 & \delta^\J_\I & B^\I_{\0\a} \\ \hline
0 & 0 & e^{-\fft12\vec c_\a\cdot\vec\phi}\,  \td\g^\a{}_\b  
\end{array}\right)\ .\label{nuh}
\ee
\medskip
Here, the indices $\a$ and $I$ label rows, while $\b$ and $J$ label
columns. 

   It follows from (\ref{newfields}) that the 3-form $F_\3$ can now be
written as 
\be
F_\3 = dA_\2 + \ft12 C_\1\, \Omega\, dC_\1\ ,\label{3form}
\ee
 where 
\be
\Omega=
\left(
\begin{array}{c|c|c}
0&0& -\oneone_n \\ \hline
0& \oneone_N & 0 \\ \hline
-\oneone_n & 0 & 0
\end{array}
\right)\ .\label{omega}
\ee
Here, $\oneone_m$ denotes the $m\times m$ unit matrix, and $n=10-D$.
The matrix $\Omega$ has eigenvalues $\pm 1$, of which $(10-D)$ are
negative and $(10-D+N)$ are positive.  In fact $\Omega$ is nothing but
a metric on the indefinite-signature flat space $\R^{10-\sst D, 10-
\sst D +\sst N}$.  It is straightforward to see that ${\cal V}$ satisfies
\be
{\cal V}^\T\, \Omega\, {\cal V} = \Omega\ .\label{nuom}
\ee
This implies that ${\cal V}$ lies in the group $O(10-D,10-D+N)$.  It follows
that $\cM = {\cal V}^\T\, {\cal V}$ satisfies the two conditions given in
(\ref{opqel}, \ref{sym}).  Furthermore, it is evident from (\ref{nuh})
that if one sets all the axions and dilatons to zero, then ${\cal V}$, and
hence $\cM$, becomes the identity.  Thus, from the discussion at the
end of subsection \ref{ssec:grassman}, it follows that ${\cal V}$ and $\cM$
give parameterisations of the coset $O(10-D,10-D+N)/(O(10-D)\times
O(10-D+N))$. (Indeed, the number of independent scalar fields in ${\cal V}$
is equal to the dimension of the coset.)  Note that $F_\3$ as given in
(\ref{3form}) is a singlet under the $O(10-D,10-D+N)$ transformations.

In principle, the proof that the $D$-dimensional Lagrangian
(\ref{ddimhet}) can be written in the form (\ref{dlagm}) could be
completed by directly evaluating $\ft14 \tr(\del\cM^{-1}\, \del \cM)$,
where $\cM={\cal V}^\T\, {\cal V}$ and ${\cal V}$ is given by
(\ref{nuh}), and showing
that it correctly reproduces the terms in the scalar sector of
(\ref{ddimhet}).  However, some more insight into the structure of the
theory can be obtained by following a slightly different approach,
showing first that ${\cal V}$ can be written as the exponential of a Lie
algebra.  Specifically, this algebra is the {\it solvable Lie
subalgebra} of the group $O(10-D,10-D+N)$.  It is to this topic that
we shall turn in the next section, where we shall be able to complete
the proof that the bosonic Lagrangians have $O(10-D,10-D+N)$ global
symmetries.

\section{Scalar cosets in the $T^n$-compactified heterotic
theory\label{sec:solvable}}

The construction of the cosets describing the scalar manifolds
arising in the toroidal reductions of eleven-dimensional
supergravity have been discussed in detail in \cite{cjlp1}.  It was
shown that the scalar-field coset in $D$-dimensional maximal
supergravity can be parametrised by the Borel subalgebra of $E_n$,
where $n=11-D$.  In other words, there is a one-to-one correspondence
between the scalar fields in the theory and the generators of the
Borel subgroup.  This shows that the scalar manifold is the coset
$E_n/K(E_n)$, where $K(E_n)$ is the maximal compact subalgebra of
$E_n$, and also shows that the group $E_n$ is of the maximally
non-compact form $E_{n(+n)}$ \cite{cjlp1}.  This latter feature is a
consequence of the fact that only the maximally non-compact form of a
group allows an Iwasawa decomposition into the product of its maximal
compact subgroup and its Borel subgroup.  Although there are other
ways to parameterise the scalar-field cosets, the Borel
parameterisation is a particularly convenient one in this context
because it is the one that arises naturally in the ``step-by-step''
dimensional reduction procedure.

In the toroidal reduction of the heterotic theory, one expects
\cite{df} the global symmetry group in $D$ dimensions to be
$O(n,n+16)\times \R$, where $D=10-n>4$.  In $D=4$ the symmetry
actually enlarges to $O(6,22)\times SL(2,\R)$
\cite{bks,deroo,ms0,sen0}, and in $D=3$ it enlarges to $O(8,24)$
\cite{ms,sen1}.  In all of these cases, the symmetry group is not
maximally non-compact, and hence a slightly different approach is
necessary in order to parameterise the relevant scalar cosets
$O(p,q)/(O(p)\times O(q))$.  This difference is reflected in the fact
that the number of scalar fields is smaller than the dimension of the
Borel subgroup of the relevant $O(p,q)$ numerator group.

     It is nonetheless convenient, in the context of dimensional
reduction, to parameterise the scalar cosets in an analogous manner.
The necessary generalisation of the Borel parameterisation is provided
by the Iwasawa decomposition \cite{iwa}. This decomposition is rather
more subtle in the case of groups that are not maximally
non-compact. One again has a unique factorisation of a group element
$g\in G$ into a product $g=k\, a\, n$, where $k$ is in the maximal
compact subgroup $K$, $a$ is in the maximal non-compact Abelian
subgroup $A$, and $n$ is in the nilpotent subgroup $N$ of $G$.  (In
the case where $G$ is maximally non-compact, $A$ is the entire Cartan
subgroup and $N$ is the strict Borel subgroup, so the product $A\, N$
belongs to the standard Borel subgroup.)  At the level of the algebra,
the Iwasawa decomposition implies that
\be
G= K\oplus G_s\ ,\label{solv}
\ee
where $K$ denotes the generators of the maximal compact subalgebra of
$G$, and $G_s$ is a so-called {\it Solvable Lie Algebra}, comprising a
subset of the Borel generators of $G$.  To be specific, it comprises
the non-compact Cartan generators $H^{\rm nc}$, together with the
subset of the positive-root generators that has strictly positive
weights under $H^{\rm nc}$.  (Clearly if $G$ were maximally
non-compact, in which case all the Cartan generators would be
non-compact, $G_s$ would comprise the entire Borel subalgebra.)

     The mathematical understanding of solvable Lie algebras relevant
to supergravity stems from Ref.\ \cite{alek}.  The application of
solvable Lie algebras has been extensively studied recently in
\cite{fre1,fre2,fre3,fre4}.  The exponential ${\cal V}=\exp(G_s)$ gives a
parameterisation of the coset $G/K$.  From this, one can construct the
$G$-invariant scalar coset Lagrangian
\bea
e^{-1}\, {\cal L}_{\rm scalar}&=&\ft14 \tr(\del_\mu \cM^{-1}\, \del^\mu\cM)
\nn\\
&=& -\ft12 \tr\Big[ \del{\cal V}\, {\cal V}^{-1}\, \Big(\del 
{\cal V}\, {\cal V}^{-1} +
(\del{\cal V}\, {\cal V}^{-1})^\#\Big)\Big]\ ,
\label{cosetlag}
\eea
where $\cM= {\cal V}^\#\, {\cal V}$. 
Here ${\cal V}^\#=\tau({\cal V}^{-1})$ where $\tau$
denotes the Cartan involution which reverses the sign of all the
non-compact generators, while leaving the sign of the compact
generators unchanged (see, for example, \cite{cjlp1}).  (For
orthogonal groups, ${\cal V}^\#$ is just equal to the transpose 
${\cal V}^\T$, and
for unitary groups it is the Hermitean conjugate ${\cal V}^\dagger$.)  The
Lagrangian is invariant under the global symmetry transformations
${\cal V}\longrightarrow {\cal V}'= {\cal O}\, {\cal V}\, \Lambda$, 
where $\Lambda$ is
any element of the group $G$, and ${\cal O}$ is a field-dependent
compensating transformation that is used to bring the transformed
coset representative ${\cal V}'$ back to the form ${\cal V}'=\exp(G_s')$.  The
Iwasawa decomposition guarantees the existence of ${\cal O}$, and the
fact that it is contained in the maximal compact subgroup $K$.  It
then follows that $\cM$ is transformed to $\cM'=\Lambda^\#\, \cM\,
\Lambda$, and hence that the Lagrangian (\ref{cosetlag}) is invariant.

     In this section, we shall show that the scalar Lagrangians for
the toroidally-reduced heterotic theory can be written in the form
(\ref{cosetlag}), where ${\cal V}\sim e^{\fft12 \vec\phi\cdot\vec H +
\chi_a\, E^a}$.  We shall obtain the explicit forms of the algebras
for the generators $\vec H$ and $E^a$, and we shall show that they are
the solvable Lie algebras associated with the global symmetry groups.
This provides an explicit derivation of the global symmetries of the
scalar sectors of the toroidally-compactified heterotic theory.  We
show that the ${\cal M}$ obtained from the scalar sector using the
Solvable Lie Algebra technique and the one obtained by studying the
coupling of the scalars with the vector potentials are equivalent,
hence completing the proof that the full Lagrangian has an $O(10-D,
10-D+N)$ global symmetry.

\subsection{$D=9$ coset}

     In $D=9$ it is easy to see how to write the scalar sector of the
Lagrangian (\ref{d9lag}) in a coset formulation.  Let us, for this
purpose, omit $\phi$, since it decouples from the rest of the scalars,
and plays no significant r\^ole in the discussion.  We introduce
generator matrices $H$ and $E_\I$, associated with the scalars
$\varphi$ and $B_\0^\I$ respectively, and we define the coset
representative
\be
{\cal V} = e^{\ft12\varphi\, H} \, e^{B_\0^\I\, E_\I}\ .\label{d9v}
\ee
 The scalar Lagrangian for $\varphi$ and $B_\0^\I$ can then be written as
\be
{\cal L} = \ft14\tr(\del_\mu\cM^{-1}\, \del^\mu \cM)\ ,\qquad
{\rm where}\qquad \cM = {\cal V}^\T\, {\cal V} \ ,\label{mdef}
\ee
 provided that the generators $H$ and $E_\I$ satisfy the algebra
\be
{[} H, E_\I {]} = \sqrt2\, E_\I\ ,\qquad {[} H, H{]} = 0 \,\qquad
{[} E_\I, E_\J {]}=0\ .\label{d9alg}
\ee
 This is a subalgebra of $O(1,N+1)$.  To see this, we first need to
establish conventions and notation for the generators and roots of the
orthogonal algebras.

    The orthogonal algebras $O(p,q)$ divide into two cases, namely
the $D_n$ series when $p+q=2n$, and the $B_n$ series when $p+q=2n+1$.
The positive roots are given in terms of an orthonormal basis $e_i$ as
follows:
\bea
D_n: && e_i\pm e_j\ , \qquad i<j \le n\ ,\nn\\
B_n: && e_i\pm e_j \ , \qquad  i<j\le n\ ,\qquad {\rm and} \qquad e_i\ ,
\label{dbroots}
\eea
where $e_i\cdot e_j=\delta_{ij}$.  It is sometimes convenient to take
$e_i$ to be the $n$-component vector $e_i=(0,0,\ldots, 0, 1, 0,
\ldots,0)$, where the ``1'' component occurs at the $i$'th position.
However, we shall find later that a different basis is more suitable
for our purposes.  The Cartan subalgebra generators, specified in a
basis-independent fashion, are $h_{e_i}$, which satisfy ${[} h_{e_i},
E_{e_j\pm e_k} {]} = (\delta_{ij} \pm \delta_{ik})\, E_{e_j\pm e_k}$,
{\it etc}.  Of these, min$(p,q)$ are non-compact, with the remainder
being compact.  It is convenient to take the non-compact ones to be
$h_{e_i}$ with $1\le i\le $min$(p,q)$.

    Returning now to our algebra (\ref{d9alg}), we find that the
generators $H$ and $E_\I$ can be expressed in terms of the
$O(1,N+1)$ basis as follows:
\bea
H &=& \sqrt2\, h_{e_1}\ ,\nn\\
E_{2k-1} &=& E_{e_1-e_{2k}} \ ,\qquad E_{2k} = E_{e_1+e_{2k}}
\qquad 1\le k\le {[}
\ft12 + \ft14 N {]}\ ,\label{d9gen}\\
E_{1+\fft12 N} &=& E_{e_1}\ ,\qquad \hbox{if $N$ is even}\ .\nn
\eea
It is easily seen that $h_{e_1}$ and $E_{e_1\pm e_i}$, together with
$E_{e_1}$ in the case of $N$ even, are precisely the generators of the
solvable Lie algebra of $O(1,N+1)$.  In other words, $h_{e_1}$ is the
non-compact Cartan generator of $O(1,N+1)$, while the other generators
in (\ref{d9gen}) are precisely the subset of positive-root $O(1,N+1)$
generators that have strictly positive weights under $h_{e_1}$.  Thus
it follows from the general discussion at the beginning of this
section that the scalar Lagrangian for the $D=9$ theory is described
by the coset\footnote{It should be emphasised that the mere fact that
one can embed the algebra (\ref{d9alg}) into the Lie algebra of a
larger Lie group $G$ does not, of itself, mean that the group $G$ acts
effectively on the scalar manifold.  Only when (\ref{d9alg}) is the
solvable Lie algebra of the group $G$ can one deduce that $G$ has an
effective group action on the scalar manifold.}
$(O(1,N+1)/O(N+1))\times \R$. (Recall that there is the additional
decoupled scalar field $\phi$ with an $\R$ shift symmetry.)

\subsection{$D=8$ coset}

   Turning now to the reduction of the heterotic theory to $D=8$, we
begin from the general toroidal reduction given in Appendix
\ref{app:hetred}, and make the following orthogonal transformation of
the dilatons:
\be
\pmatrix{\phi_1\cr \phi_2\cr \phi_3} = 
 \pmatrix{-\sqrt{\ft34} & \sqrt{\ft18} & \sqrt{\ft18} \cr
 \sqrt{\ft3{28}} & -\sqrt{\ft1{56}} & \sqrt{\ft78} \cr
 \sqrt{\ft17} & \sqrt{\ft67} & 0}\, 
\pmatrix{\phi \cr \varphi_1\cr \varphi_2}\ .\label{d8redef}
\ee
In terms of these rotated fields, the Lagrangian for the scalar
subsector of the eight-dimensional theory becomes
\bea
e^{-1}\, {\cal L}_8 &=& -\ft12(\del\phi)^2 -\ft12(\del\varphi_1)^2
-\ft12(\del\varphi_2)^2 -\ft12 e^{\sqrt2(\vp_1+\vp_2)}\, (\del
A_{\023} + B^\I_{\0 2}\, \del B^\I_{\0 3})^2 \label{d8scal}\\
&&-\ft12 e^{\sqrt2(\vp_1-\vp_2)}\, (\del \cA^2_{\0 3})^2 -\ft12
e^{\sqrt2\vp_2}\, (\del B^\I_{\0 2})^2 -\ft12 e^{\sqrt2\vp_1}\, 
(\del B^\I_{\0 3} - \cA^2_{\0 3}\, \del B^\I_{\0 2})^2\ .\nn
\eea
For generality, we again allow the range of the index $I$ to be $1\le
I\le N$, rather than just the specific range $1\le I\le 16$ that
arises in the heterotic theory. The eight-dimensional string coupling
constant is given by $\lambda_8 = e^{\sqrt{3/4}\, \phi}$.  Note that
the dilaton $\phi$ decouples from the rest of the scalars.  We shall
therefore temporarily suppress $\phi$ in the following discussion of
the coset structure of the scalar manifold, with the understanding
that its constant shift symmetry contributes an additional independent
$\R$ factor to the full global symmetry.

We can then show that the Lagrangian (\ref{d8scal}), with $\phi$
omitted, can be obtained by parameterising a coset as
\be
{\cal V} = e^{\ft12 \vec\varphi\cdot\vec H}\, e^{\cA^2_{\03}\, E_2{}^3}\, 
e^{A_{\0 23} \, V^{23}}\,
e^{B^\I_{\0 2}\, U^2_\I}\, e^{B^\I_{\0 3}\, U^3_\I}\ ,\label{d8nu}
\ee
 and substituting this into the first line of (\ref{cosetlag}),
with $\cM= {\cal V}^\T\, {\cal V}$.  The commutation relations for the various
generators can then be read off by noting that the 1-form field
strengths are given by \cite{cjlp1}
\be
{\cal G} = d{\cal V}\, {\cal V}^{-1}= \ft12 d\vec\varphi\cdot\vec H + \cF^2_{\1
3}\, E_2{}^3 + F_{\1 23}\, V^{23} + G^\I_{\1 2}\, U^2_\I 
+ G^\I_{\1 3}\, U^3_\I \ .\label{d8dv}
\ee
Comparing with the explicit expressions given in Appendix
\ref{app:hetred} and in (\ref{d8scal}), we find that
\bea
&&{[} H_1, V^{23} {]} = \sqrt2\, V^{23}\ ,
\qquad {[} H_2, V^{23} {]} = \sqrt2\, V^{23}\ ,\nn\\
&&{[} H_1, E_2{}^3 {]} = \sqrt2\, E_2{}^3\ ,
\qquad {[} H_2, E_2{}^3 {]} = -\sqrt2\, E_2{}^3\ ,\nn\\
&&{[} H_1, U^2_\I {]} = 0\ ,\qquad {[} H_2, U^2_\I {]} = 
\sqrt2\, U^2_\I\ ,\nn\\
&&{[} H_1, U^3_\I {]} = \sqrt2\, U^3_\I\ ,
\qquad {[} H_2, U^3_\I {]} = 0\ ,\nn\\
&&{[} U^2_\I, U^3_\J {]} =  \delta_{\I\J}\, V^{23}\ ,\qquad
{[} E_2{}^3, U^2_\I {]} = - \, U^3_\I\ ,\label{d8comm}
\eea
with all other commutators vanishing.  

     We shall now show that the algebra (\ref{d8comm}) is precisely
the solvable Lie algebra for $O(2,N+2)$ (or, in other words, that the
exponentiation of (\ref{d8comm}) gives a parameterisation of the coset
$O(2,N+2)/(O(2)\times O(N+2))$). To do this, it is instructive to look
first at two examples, namely $O(2,3)$ and $O(2,4)$ corresponding to
$N=1$ and $N=2$.  From (\ref{dbroots}), the positive roots for
$O(2,3)$ are $e_1-e_2$, $e_1+e_2$, $e_1$ and $e_2$.  On the other
hand, from (\ref{d8comm}) we have four positive-root generators in
this case, namely $E_2{}^3$, $V^{23}$, $U^2_1$ and $U^3_1$.  It is
easy to see that the commutation relations in (\ref{d8comm}) lead to
the identifications
\bea
&&E_2{}^3 = E_{e_1-e_2}\ ,\qquad V^{23} = E_{e_1+e_2}\ ,\qquad
U^2_1 = E_{e_2}\ ,\qquad U^2_1 = E_{e_1}\ ,\nn\\
&&H_1 = \sqrt 2\, h_{e_1}\ ,\qquad H_2 = h_{e_2}\ .\label{o23cor}
\eea
Note that in this case, because $O(2,3)$ is maximally non-compact, all
of the Borel generators of $O(2,3)$ occur in the associated solvable
Lie algebra.

    For the case $O(2,4)$, we have generators $E_2{}^3$, $V^{23}$,
$U^2_\I$ and $U^3_\I$ in our coset parameterisation (\ref{d8nu}),
where $1\le I \le 2$.  From (\ref{dbroots}), the positive roots of
$O(2,4)$ are $e_1\pm e_2$, $e_1\pm e_3$ and $e_2\pm e_3$.  From the
algebra (\ref{d8comm}), it is clear that we should take 
\be
H_1=\sqrt2\, h_{e_1}\\ ,\qquad H_2 = \sqrt2\, h_{e_2} \, \qquad
E_2{}^3 = E_{e_1-e_2}\ ,\qquad V^{23} = E_{e_1+e_2}\ .
\ee
 It is then evident that in order for the remaining generators to have
the proper weights under $H_1$ and $H_2$, we must have
\bea
&&U_1^2 = \a_1\, E_{e_2+e_3} + \b_1\, E_{e_2-e_3}\ ,\qquad
U_1^3 = \a_1\, E_{e_1+e_3} + \b_1\, E_{e_1-e_3}\ ,\nn\\
&&U_2^2 = \a_2\, E_{e_2+e_3} + \b_2\, E_{e_2-e_3}\ ,\qquad
U_2^3 = \a_2\, E_{e_1+e_3} + \b_2\, E_{e_1-e_3}\ ,\label{o24rel}
\eea
 for appropriate constants $\a_1$, $\b_1$, $\a_2$ and $\b_2$.  Choosing
our sign conventions for $O(2,4)$ so that
\be
{[} E_{e_1-e_2} , E_{e_2\pm e_3} {]} = -E_{e_1\pm e_3}\ ,\qquad
{[} E_{e_1\pm e_3}, E_{e_2\mp e_3} {]} = - E_{e_1+e_2}\ ,
\ee
 and bearing in mind that the relations (\ref{o24rel}) should preserve
the strengths of the generators, we find that, up to arbitrariness in
the phases, the solution is $\a_1=\b_1=1/\sqrt2$,
$\a_2=-\b_2=\im/\sqrt2$.  Thus we have
\bea
&&U_1^2 = \ft1{\sqrt2}\,( E_{e_2+e_3} + E_{e_2-e_3})\ ,\qquad
U_1^3 = \ft1{\sqrt2}\, (E_{e_1+e_3} + E_{e_1-e_3})\ ,\nn\\
&&U_2^2 = \ft{\im}{\sqrt2}\, (E_{e_2+e_3} - E_{e_2-e_3})\ ,\qquad
U_2^3 = \ft\im{\sqrt2}\, (E_{e_1+e_3} - E_{e_1-e_3})\ ,\label{o24rel2}
\eea

    It is now straightforward to generalise the result to the generic
case $O(2,N+2)$. The new feature that we have seen for $N=2$, where
the generators for the Yang-Mills axions are expressed in terms of
certain real or imaginary combinations of the $O(2,4)$ generators,
persists for all $N\ge2$.  We find that the embedding of the
generators in (\ref{d8nu}) into $O(2,N+2)$ is as follows.  When $N$ is
even, we find
\bea
&& H_1 = \sqrt2\, h_{e_1}\ ,\qquad H_2 = \sqrt2\, h_{e_2}\ ,\qquad
  E_2{}^3 = E_{e_1-e_2} \ ,\qquad V^{23} = E_{e_1+e_2}\ ,\nn\\
&&U^2_{2k-1} = \ft1{\sqrt2}\, (E_{e_2+e_{k+2}} + E_{e_2-e_{k+2}}) \ ,\qquad
U^3_{2k-1} = \ft1{\sqrt2}\, (E_{e_1+e_{k+2}} + E_{e_1-e_{k+2}})\ ,\nn\\
&&U^2_{2k} = \ft\im{\sqrt2}\, (E_{e_2+e_{k+2}} - E_{e_2-e_{k+2}}) \ ,\qquad
U^3_{2k} = \ft\im{\sqrt2}\, (E_{e_1+e_{k+2}} - E_{e_1-e_{k+2}})
\ ,\label{o22n}
\eea
 where $k$ has the range $1\le k \le [\ft12 N]$.  If $N$ is odd, in
addition to the identifications (\ref{o22n}) for $1\le k\le {[} \ft12
N{]}$, we have 
\be
U^2_{\sst N} = E_{e_2}\ ,\qquad U^3_{\sst N} = E_{e_1}\ .
\ee
 (This embedding of the generators of the solvable Lie algebra in
$O(2,N+2)$ was also encountered in \cite{fre3}.)

     It is easy to see that the subset of $O(2,N+2)$ generators
$h_{e_1}$, $h_{e_2}$, $E_{e_1\pm e_2}$, $E_{e_1\pm e_i}$ and
$E_{e_2\pm e_i}$, with $3\le i \le 2+{[}\ft12 N{]}$, together with
$E_{e_1}$ and $E_{e_2}$ if $N$ is odd, precisely constitute the set of
generators in the solvable Lie algebra of $O(2,N+2)$.  This is because
$h_{e_1}$ and $h_{e_2}$ are the two non-compact Cartan generators of
$O(2,N+2)$, and the positive-root generators we have just listed are
the full set that have strictly positive weights under $h_{e_1}$ and
$h_{e_2}$.  Thus it follows that the quantity ${\cal V}$ defined in
(\ref{d8nu}) gives a parameterisation of the coset
$O(2,N+2)/(O(2)\times O(N+2))$.  Together with the shift symmetry of
the dilaton $\phi$, this shows that the scalar Lagrangian
(\ref{d8scal}) is invariant under global $O(2,N+2)\times \R$
transformations.

\subsection{Cosets in $D\ge5$\label{ssec:dge5}}

    Having seen how the coset construction works in the special cases
in $D=9$ and $D=8$, we are now in a position to consider the general
$D$-dimensional case.  However, owing to the fact that higher-degree
fields can be dualised to give additional scalars in $D=4$ and $D=3$,
we shall treat these two dimensions separately, having first
considered the more straightforward cases $D\ge5$.

   From Appendix \ref{app:hetred}, the scalar Lagrangian in $D$
dimensions can be expressed as
\be
e^{-1}\, {\cal L}_{\sst D} 
  =-\ft12 (\del\vec\vp)^2 -\ft12 \sum_{i<j} e^{\vec
b_{ij}\cdot \vec\vp}\, ({\cal F}^i_{\0 j})^2 - \ft12 \sum_{i<j}
e^{\vec a_{ij}\cdot\vec\vp}\, (F_{\1 ij})^2  -\ft12 \sum_{i,\I}
e^{\vec c_i\cdot\vec\vp}\, (G^\I_{\1 i})^2 \ ,\label{dlag}
\ee
together with a free Lagrangian for the dilaton
\be \phi= -\sqrt{\ft{D-2}{8}}\, \vec a_{123}\cdot\vec\phi\ , \ee which
is decoupled from (\ref{dlag}).  The 3-form field strength $F_\3$
couples only to $\phi$, and the string coupling constant is given by
$\Lambda_{\sst D} = \exp(\sqrt{(D-2)/8}\, \phi)$.  We shall, as usual,
concentrate only on the sector with $\phi$ omitted during our
discussion of the global symmetries. Note that here the notation for
the dilaton vectors used here is a little different from the one
introduced in Appendix \ref{app:hetred}. Since the dilaton $\phi$ has
been truncated out, the dilaton vectors in (\ref{dlag}) have $(10-D)$
components rather than $(11-D)$.  They are given by
\be
\vec b_{ij} = \sqrt2(-\vec e_i + \vec e_j)\ ,
 \qquad \vec a_{ij} = \sqrt2(\vec e_i + \vec e_j)\ ,\qquad
\vec c_i = \sqrt2\, \vec e_i\ .\label{dvs}
\ee
We have also changed from indices $\a,\b,\ldots$ which range from 2 to
$(11-D)$ to $i,j,\ldots$ which range from 1 to $(10-D)$.  Since there
will be no confusion, we shall use the same symbols $\vec b_{ij}$ and
$\vec c_i$ as in Appendix \ref{app:hetred}, and $\vec a_{ij}$ in place
of $\vec a_{1\a\b}$.  The 1-form field strengths in (\ref{dlag}) are
given by
\bea
{\cal F}^i_{\1 j} &=& \gamma^k{}_j\, d{\cal A}^i_{\0 k}\ ,\nn\\
F_{\1 ij} &=& \gamma^k{}_i\, \gamma^\ell{}_j\, (dA_{\0 k\ell} +
B^\I_{\0 [k}\, dB^\I_{\0 \ell ]})\ ,\label{dfield}\\
G^a_{\1 i} &=& \gamma^j{}_i\, dB^\I_{\0 j}\ .\nn
\eea

   We find that we can write the Lagrangian (\ref{dlag}) in the form
(\ref{cosetlag}), where the coset representative ${\cal V}$ is parametrised
as \cite{cjlp1}
\be
{\cal V} = e^{\ft12\vec\phi\cdot\vec H}\, e^{{\cal A}^i_{\0 j}\, E_i{}^j}\,
   e^{\fft12 A_{\0 ij}\, V^{ij}}\, e^{B^\I_{\0 i}\, U_\I{}^i}
\ .\label{dnu}
\ee
The commutation relations for the various generators can be determined
by comparing the expression for the field strengths 
\be
d{\cal V}\, {\cal V}^{-1} = \ft12 d\vec\phi\cdot\vec H + \sum_{i<j} 
e^{\fft12 \vec
b_{ij}\cdot\vec \phi}\, {\cal F}^i_{\1 j}\, E_i{}^j +
\sum_{i<j} e^{\fft12 \vec
a_{ij}\cdot\vec \phi}\, F_{\1 ij} \, V^{ij} + \sum_{i,\I} e^{\fft12 \vec
c_i\cdot\vec\phi}\, G^\I_{\1 i}\, U_\I{}^i \label{want}
\ee
 with the expressions given in (\ref{dfield}).  We find that the
non-vanishing commutators are given by
\bea
&&{[} \vec H, E_i{}^j {]} = \vec b_{ij}\, E_i{}^j\ ,\qquad
{[}\vec H, V^{ij} {]} = \vec a_{ij}\, V^{ij}\ ,\qquad
{[}\vec H, U_\I{}^i {]} = \vec c_i\, U_\I{}^i\ ,\nn\\
&&{[} E_i{}^j, E_k{}^\ell {]} =\delta^j_k\, E_i{}^\ell
-\delta_i^\ell\, E_k{}^j\ ,\nn\\
&&{[} E_i{}^j, V^{k\ell} {]} = -\delta_i^k\, V^{j\ell} -
\delta_i^\ell\, V^{kj}\ ,\qquad
{[} E_i{}^j, U_\I{}^k {]} = - \delta_i^k\, U_\I{}^j\ ,\nn\\
&&{[}U_\I{}^i, U_\J{}^j {]} = \delta_{\I\J}\, V^{ij}\ .\label{dcomm}
\eea
 The way in which the multiple commutators arising in the evaluation
of $d{\cal V}\, {\cal V}^{-1}$ conspire to produce the precise expressions
(\ref{dfield}) is discussed in detail in \cite{cjlp1,cjlp2}.  

    We shall now show that the above set of generators and their
commutation relations can be embedded into those of $O(10-D,10-D+N)$,
and that in fact they precisely correspond to the solvable lie algebra
of $O(10-D,10-D+N)$.  To see this, it is useful first to introduce the
set of orthonormal vectors $\td e_i$, related to $e_i$ by
\be
\td e_i = e_{11-\sst{D}-i}\ , \qquad 1\le i\le 10-D\ .\label{redefe}
\ee
 Let us, for definiteness, first consider the case where $N$ is even.
We then find that the above generators can be written in terms of
those of $O(10-D,10-D+N)$ as follows.  The generators $\vec H$, 
$E_i{}^j$ and $V^{ij}$ are written as
\be
H_i= \sqrt2\, h_{\td e_i}\ ,\qquad 
E_i{}^j = E_{-\td e_i+ \td e_j}\ ,\qquad V^{ij} = E_{\td e_i+\td e_j}\ ,
\ee
where we have $1\le i < j \le 10-D$.  For the generators $U_\I{}^i$
associated with the Yang-Mills axions, we find that we can write
\bea
U^i_{2k-1} &=& \ft1{\sqrt2}\, (E_{\td e_i +e_{k+m}} +E_{\td
  e_i-e_{k+m}}) \ ,\nn\\
U^i_{2k} &=& \ft\im{\sqrt2}\, (E_{\td e_i +e_{k+m}} -E_{\td
  e_i-e_{k+m}}) \ , \label{dmd}
\eea
 where $m=10-D$ and $1\le k\le [\ft12 N]$. If $N$ is odd, then in
addition we have
\be
U^i_{\sst N} = E_{\td e_i}\ .
\ee

It is easily seen that this set of generators comprises the solvable
Lie algebra of $O(10-D,10-D+N)$.  In other words, they are written in
terms of the complete set of non-compact Cartan generators of
$O(10-D,10-D+N)$, together with all the positive-root generators that
have strictly positive weights under the non-compact Cartan
generators.

     So far in this subsection, we have constructed the coset
representative ${\cal V}$ for the purpose of writing the scalar Lagrangian
in the form $\ft14\tr(\del \cM^{-1}\, \del \cM)$, where $\cM={\cal V}^\T\,
{\cal V}$.  Using the solvable Lie algebra formalism, we have accordingly
shown that the scalar Lagrangian is described by the coset
$O(10-D,10-D+N)/((O(10-D)\times O(10-D+N))$ (together with an extra
$\R$ factor for the scalar field $\phi$).  This construction is
abstract, in the sense that we have not taken any specific realisation
for the generators; they are simply required to satisfy (\ref{dcomm}).
On the other hand, in section \ref{sec:globsym} we have also obtained
an expression for a coset parameterisation ${\cal V}$, by considering the
coupling of the scalars to the 1-form potentials.  Since these
potentials form a fundamental representation of $O(10-D,10-D+N)$, the
representation for ${\cal V}$ that we obtained there was necessarily given
in terms of matrices of dimension $(20-2D+N)$.  To complete the proof
that the entire $D$-dimensional Lagrangian has a global
$O(10-D,10-D+N)$ symmetry, we need to make contact between the two
descriptions, by showing explicitly that we can write ${\cal V}$ as given by
(\ref{nuh}) in the form (\ref{dnu}), and by showing that the generator
matrices satisfy the commutation relations (\ref{dcomm}).

    To do this, we need only look at the form of (\ref{nuh}) in the
neighbourhood of the identity, in which case it is easy to read off
the generator matrices associated with each of the scalar
fields.\footnote{It is necessary, once again, to translate between the
$2\le\a\le 11-D$ notation and the $1\le i \le 10-D$ notation.}  By
this means, we see that the generators are given as follows:
\bea
&&\vec H_i = 
\left(\begin{array}{c|c|c}
       \sum_i \vec c_i\, e_{ii} & 0 & 0 \\ \hline
        0 & 0 & 0 \\ \hline
        0 & 0 & -\sum_i \vec c_i\, e_{ii}
        \end{array}\right) \ ,\qquad
E_i{}^j = \left(\begin{array}{c|c|c}
                 -e_{ji} & 0 & 0\\ \hline
                        0 & 0 & 0\\ \hline
                        0 & 0 & e_{ij} 
         \end{array}\right)\ ,\nn\\
&&V^{ij} = \left(\begin{array}{c|c|c}
                  0 & 0 &  e_{ij}- e_{ji} \\ \hline
                        0 & 0 & 0 \\ \hline
                        0 & 0 & 0 
         \end{array}\right)\ ,\qquad
U^i_\I = \left(\begin{array}{c|c|c}
                  0 & e_{i\I} &  0 \\ \hline
                        0 & 0 & e_{\I i} \\ \hline
                        0 & 0 & 0 
         \end{array}\right)\ . \label{generators}
\eea
Here, each $e_{ab}$ is defined to be a matrix of the appropriate
dimensions that has zeroes in all its entries except for a 1 in the
entry at row $a$ and column $b$.  These satisfy the matrix product
rule $e_{ab}\, e_{cd} = \delta_{bc}\, e_{ad}$.  It is not hard to show
that these matrices indeed satisfy the commutation relations
(\ref{dcomm}).  This completes our demonstration that the entire
Lagrangian has a global $O(10-D,10-D+N)$ symmetry.

\subsection{$D=4$ coset}

   In four dimensions there is an additional axion, over and above
those of the generic $D$-dimensional discussion, which arises if the
2-form potential $A_\2$ is dualised.  If $A_\2$ is left undualised,
the scalar Lagrangian will have an $O(6,N+6)\times \R$ global
symmetry, as one would expect from the general results of the previous
subsection.  If $A_\2$ is dualised, the symmetry group enlarges to
$O(6,6+N)\times SL(2,\R)$.  We can see this very easily in the
formalism that we have been using in this paper.

    To include the effect of dualising $A_\2$ to give an additional
axion, we first add the kinetic term for $A_\2$ to the scalar
Lagrangian.  Together with the kinetic term for $\phi$, this extra
term gives
\be
e^{-1}\, {\cal L}_{\rm extra} = -\ft12(\del\phi)^2 
-\ft12 e^{-2\phi}\, F_\3^2\ ,\label{d4extra}
\ee
where $\phi=-\ft12\vec a_{123}\cdot\vec\phi$.  This is the linear
combination of the dilatons which, as discussed in the previous
subsection, is decoupled from the rest of the scalar Lagrangian.  In
the absence of the extra term (\ref{d4extra}), it would be responsible
for contributing the extra $\R$ factor in the global symmetry.  If we
now dualise $A_\2$, the term (\ref{d4extra}) gives the additional
contribution
\be
e^{-1}\, {\cal L}_{(\phi,\chi)} = 
-\ft12(\del\phi)^2 -\ft12 e^{2\phi}\, (\del\chi)^2 \label{duald4extra}
\ee
 to the scalar Lagrangian, where
\be
F_\3 = e^{-2\phi}\, {*d\chi}\ ,
\ee
and $\chi$ is the new axion dual to $A_\2$.  Since the dilaton/axion
system $(\phi,\chi)$ is decoupled from the rest of the scalar
Lagrangian, it follows that the total global symmetry group for the
scalar sector is now the direct product $O(6,6+N)\times SL(2,\R)$.

   This global symmetry extends to the full four-dimensional theory.
To see this, we note that since the Bianchi identity for $F_\3$ gives
$dF_\3 = \ft12 dB_\1^\I\wedge dB_\1^\I - dA_{\1\a}\wedge d\cA_\1^\a$,
the dualisation of $A_\2$ to $\chi$ will also give the extra
contribution $\ft12 \chi\, {*(} dB_\1^\I\wedge dB_\1^\I -2
dA_{\1\a}\wedge d\cA_\1^\a)$ to the dualised Lagrangian.  In the
notation of section \ref{ssec:globalsym}, the full Lagrangian can
therefore be written as
\bea
e^{-1}\, {\cal L}_4 &=& R - \ft12(\del\phi)^2 -\ft12 e^{2\phi}\,
(\del\chi)^2 + \ft14\tr(\del\cM^{-1}\, \del \cM)\ ,\nn\\ 
&&-\ft14 e^{-\phi}\, H_\2^\T\, \cM\, H_\2 +\ft12\, \chi\, {*(} H_\2^\T\,
\Omega\, H_\2)\ ,\label{4lagm}
\eea
where $\cM$ is a parameterisation of the coset $O(6,N+6)/(O(6)\times
O(N+6))$.  As well as the manifest global $O(6,N+6)$ symmetry of the
Lagrangian, there is also an $SL(2,\R)$ symmetry of the equations of
motion, under which ${\cal V}\, H_\2$ and $e^{-\phi}\, 
{*({\cal V}\, H_\2)^\T}$ form an $SL(2,\R)$ doublet.

\subsection{$D=3$ coset \label{ssec:d3coset}}

    In three dimensions, the discussion of subsection \ref{ssec:dge5}
shows that if one leaves the higher-degree fields in their undualised
form, the global symmetry group will be $O(7,7+N)\times \R$. If one
dualises the vector potentials $(\cA_\1^\a, A_{\1 \a}, B_\1^\I)$ to give
an additional $7+7+N$ axions $(\wtd\chi_\a, \chi^\a, \lambda_\I)$, then
the global symmetry group enlarges to $O(8,8+N)$.  Note that the
entire bosonic sector is now composed only of scalar fields.

   To see how the symmetry enlarges, we first perform the dualisations
specified above.  To do this, we begin by obtaining the Bianchi
identities for the 2-form field strengths.  From the results in
Appendix \ref{app:hetred}, we find the following:
\bea
d\cF_\2^\a &=& \cF^\a_{\1 \b} \wedge \cF_\2^\b\ ,
\nn\\
dF_{\2 \a} &=& - \cF^\b_{\1 \a}\wedge F_{\2 \b} - F_{\1\a\b}
\wedge \cF_\2^\b
+ G^\I_{\1 \a}\wedge G^\I_\2\ ,\label{d3bianchi}\\
dG^\I_\2 &=& G^\I_{\1 \a}\wedge \cF_\2^\a\ .\nn
\eea
 Adding the appropriate Lagrange multipliers to the original
Lagrangian, namely the terms
\bea
{\cal L}_{\rm LM} &=& \wtd\chi_\a\, (d\cF_\2^\a -\cF^\a_{\1 \b} \wedge
\cF_\2^\b) +\lambda_\I\, (dG^\I_\2 - G^\I_{\1 \a}\wedge \cF_\2^i\a) \nn\\
&&-\chi^\a\, (dF_{\2 \a} + \cF^\b_{\1 \a}\wedge F_{\2 \b} +F_{\1\a\b}
\wedge \cF_\2^\b - G^\I_{\1 \a}\wedge G^\I_\2)\ ,
\eea
 we find, after eliminating the 2-form field strengths by solving
algebraically for them, that the dualised 1-form field strengths are
given by
\bea
\cF_{\1 \a} &\equiv& - e^{\vec b_\a\cdot\vec\phi}\, {*\cF_\2^\a} =
  d\wtd\chi_\a + \wtd\chi_\b\, \cF^\b_{\1\a} - \chi^\b\, F_{\1\a\b} +
\lambda_\I\, G^\I_{\1\a} \ ,\nn\\
F^\a_{\1}&\equiv& e^{\vec a_{1\a}\cdot\vec\phi}\, {*F_{\2\a}} =
 d\chi^\a - \chi^\b\, \cF^\a_{\1\b} \ ,\label{d3dualised}\\
G_{\1\I} &\equiv& - e^{\vec c\cdot\vec\phi}\, {*G^\I_\2} = d\lambda_\I -
\chi^\a\, G^\I_{\1\a}\ .\nn
\eea
(The sign differences in the Lagrange multiplier and definition of the
field strength for the $\chi^\a$ terms are purely conventional, and
have been introduced in order to simplify the form of the final result.)

     Looking at the dilaton vectors for the full set of 1-form field
strengths, we find that they are as follows:
\bea
{\cal F}^\a_{\1 \b}: && \vec b_{\a\b} = -\vec c_\a + \vec c_\b\ ,\qquad
F^\a_\1:\qquad -\vec a_{1\a}= -\vec c_\a + \vec c_9\ ,\nn\\
F_{\1 \a\b}: && \vec a_{1\a\b} = \vec c_\a + \vec c_\b\ ,\qquad
{\cal F}_{\1 \a}:\qquad -\vec b_\a=\vec c_\a + \vec c_9\ ,\label{pairs}\\
G^\I_{\1 \a}:&& \vec c_\a\ ,\qquad \qquad\qquad \qquad
G_{\1 \I}:\qquad -\vec c = \vec c_9\ ,\nn
\eea
 where 
\be
\vec c_\sA = (\vec c_\a,\vec c_9)\ ,\qquad 
\vec c_{\sA}\cdot\vec c_{\sB}=2\, \delta_{\sA\sB}\ .
\ee
 We see that by writing the dilaton vector $-\vec c$ for the axions
coming from the dualisation of $G^\I_\2$ as $\vec c_9$, it is then
natural to extend the index range from $\vec c_\a$ with $2\le \a\le 8$
to $\vec c_\sA$ with $2\le A\le 9$.  The sets of field strengths on
each line in (\ref{pairs}) then naturally pair together to make an
extended set.

    Let us define new extended sets of potentials ${\wb A}_{\0\sA\sA}$,
${\wb \cA}^\sA_{\0\sB}$ and ${\wb B}^\I_{\0\sA}$ by
\bea
&&{\wb A}_{\0 \a\b} = A_{\0\a\b}\ ,\qquad {\wb A}_{\0 \a 9} =
\td\gamma^\b{}_\a\, \wtd\chi_\b + \ft12 B^\I_{\0 \a}\, \lambda_\I\ ,\nn\\
&&{\wb \cA}^\a{}_{\0\b} = \cA^\a{}_{\0\b}\ ,\qquad
{\wb \cA}^\a_{\0 9} = \chi^\a\ ,\label{extended}\\
&&{\wb B}^\I_{\0\a} = B^\I_{\0\a} \ ,\qquad
{\wb B}^\I_{\0 9} = \lambda_\I \ .\nn
\eea
 We also can define an extended set of matrices ${\wb\gamma}^\sA{}_\sB$,
and their inverses $\wtd{\wb\gamma}^\sA{}_\sB\equiv \delta^\sA_\sB +
{\wb \cA}^\sA_{\0\sB}$.  From the definitions (\ref{extended}), it
follows that
\be
{\wb \gamma}^\a{}_\b = \gamma^\a{}_\b\ ,\qquad
{\wb \gamma}^\a{}_9 = \gamma^\a{}_\b\, \chi^\b\ ,\qquad
{\wb \gamma}^9{}_9 = 1\ .
\ee
 (As usual, ${\wb\gamma}^\sA{}_\sB$ is zero if $A>B$.)  From the
above, we then find that the extended set of 1-form field strengths
${\wb F}_{\1\sA\sA}$, ${\wb \cF}^\sA_{\1\sB}$ and ${\wb G}^\I_{\1\sA}$
can be written as
\bea
{\wb F}_{\1\sA\sB}&=& {\wb \g}^{\sC}{}_\sA\, {\wb \g}^\sD{}_\sB \, 
(d{\wb A}_{\0 \sC\sD} - {\wb B}^\I_{\0[\sC}\, d{\wb B}^\I_{\0\sD]})
\ ,\nn\\
{\wb \cF}^{\sA}_{\1\sB} &=& {\wb \g}^\sC{}_\sB\, 
d{\wb \cA}^\sA{}_{\0\sC}\ ,\qquad
{\wb G}^\I_{\1 \sA} = {\wb \g}^\sB{}_\sA\, 
d{\wb B}^\I_{\0\sB}\ .
\eea

    In terms of these extended sets of fields, the fully-dualised
three-dimensional Lagrangian is given by
\be
e^{-1}\, {\cal L}_3 
  =-\ft12 (\del\vec\phi)^2 -\ft12 \sum_{\sA<\sB} e^{\vec
b_{\sA\sB}\cdot \vec\phi}\, ({\wb \cF}^\sA_{\0 \sB})^2 
- \ft12 \sum_{\sA<\sB}
e^{\vec a_{\sA\sB}\cdot\vec\phi}\, ({\wb F}_{\1 \sA\sB})^2  
-\ft12 \sum_{\sA,\I}
e^{\vec c_\sA\cdot\vec\phi}\, ({\wb G}^\I_{\1 \sA})^2 
\ ,\label{d3lag}
\ee
 where $\vec b_{\sA\sB} = -\vec c_\sA + \vec c_\sB$, $\vec a_{\sA\sB}
= \vec c_{\sA} + \vec c_{\sB}$ and $\vec c_{\sA}\cdot \vec c_\sB
=2\delta_{\sA\sB}$.  Thus we see that the $D=3$ Lagrangian has the
identical form as (\ref{dlag}), except that the index range of $A$ is
extended to $2\le A\le 9$, rather than the $1\le i\le 10-D$ range
occurring in (\ref{dlag}).  It follows from the discussion in
subsection \ref{ssec:dge5} that the Lagrangian (\ref{d3lag}) is
therefore described by the coset $O(8,8+N)/(O(8) \times O(8+N))$.

\section{Time-like reductions of the heterotic string theory}

    Dimensional reductions on Lorentzian tori have been discussed in
\cite{moore,gpr,stelle,hj,clpsst,hull1}.  It was shown that the global
symmetry groups remain unchanged from those of the usual
Euclidean-torus reductions, but the coset structure is changed by
virtue of the fact that the previous compact denominator groups are
replaced by certain non-compact versions of these groups.

    In this section, we give explicit derivations of the coset
structure in all dimensions $3\le D\le 9$, using the techniques that
we have presented earlier in this paper.  As we showed in section
\ref{ssec:grassman}, the denominator groups $K$ for cosets $G/K$,
where $G=O(p,q)$, are determined by the choice of a fiducial matrix
$W_0$ lying on a particular orbit of the matrices $W$ satisfying
(\ref{opqel}, \ref{sym}).  In the discussion in section
\ref{ssec:globalsym}, we saw that the matrices $\cM$ that parameterise
the scalar manifolds in dimensionally reduced Lagrangians played the
r\^ole of the matrices $W$, and that the fiducial matrix $W_0$ could
be read off simply by setting all the scalar fields to zero.  In the
usual reductions on Euclidean tori, the fiducial matrix is always just
the identity, and hence it follows that the denominator group is the
compact form $O(p)\times O(q)$.

     For reductions on Lorentzian tori, a general discussion can
easily be given for the cases $D\ge5$.  As in section
\ref{ssec:globalsym}, the $\cM$ matrix can be read off from the
kinetic terms of the 2-form field strengths.  The fiducial matrix
where all the scalars vanish is diagonal, with unit-magnitude
components whose signs are determined by the signs of the kinetic
terms for the corresponding 2-form field strengths.  Without loss of
generality,\footnote{It was shown in \cite{clpsst} that time-like and
space-like reduction steps commute.}  let us consider the case where
the time-like reduction step is the one from $D=10$ to $D=9$,
corresponding, in our notation, to the internal index $\a$ taking the
value 2. It follows that all lower-dimensional fields that have a
single $\a=2$ internal index suffer a sign-reversal for their kinetic
terms \cite{clpsst}.  This means that the two 2-form field strengths
$\cF_\2^\a$ and $F_{\2\a}$ with $\a=2$ will acquire sign-reversed
kinetic terms, while all the other 2-forms will retain their standard
signs.  This pair of 2-forms can be seen to be associated with one
symmetric pair of off-diagonal $-1$ entries in the metric $\Omega$
given in (\ref{omega}), and thus they are associated with one
eigenvalue $+1$ and one eigenvalue $-1$ in the
$O(10-D,10-D+N)$-invariant metric $\Omega$.  This means that in terms
of the diagonal invariant metric that we used in section
\ref{ssec:grassman}, the fiducial matrix $W_0$ is given by
(\ref{fiducial}) with $m=n=1$.  Thus the coset spaces describing
scalar manifolds in the $D$-dimensional theories obtained by time-like
reductions to $D\ge5$ are
\be
\fft{O(10-D,10-D+N)}{O(1,9-D)\times O(1,9-D+N)}\, \times \R\
.\label{d5time}
\ee

There is an alternative way to determine the fiducial matrix $W_0$,
which will prove to be useful later when we look at the time-like
reduction down to $D=3$.  From the form of the Kaluza-Klein metric
reduction ansatz (\ref{dmetric}), we see that the effect of making a
time-like reduction in the step from $D=10$ to $D=9$ can be achieved
by performing the complex field redefinition
\be
\vec\phi\longrightarrow \vec\phi + \ft{\im\, \pi}{2}\, (\vec c_2 -\vec
c)\ .\label{complex}
\ee
 (Note that the ten-dimensional dilaton $\phi_1$ is unchanged under
this redefinition.)  From (\ref{nuh}), we see that upon setting all the
axions to zero, the matrix $\cM={\cal V}^\T\, {\cal V}$ is given by
\be
\cM= {\rm diag}\, (e^{\vec c_\a\cdot\vec\phi}\, \delta^\b_\a\, ,
    \, \delta^\J_\I \, , \, 
          e^{-\vec c_\a\cdot\vec\phi}\, \delta^\a_\b)\ ,\label{cmu}
\ee
 for the usual case of reduction on a Euclidean torus.  The
transformation (\ref{complex}) then implies that there is a sign
reversal on the two components corresponding to $\a=2$.  If we now set
the dilatons to zero, we get precisely the same fiducial matrix as
described above.  

     In $D=4$, the global symmetry group is $O(6,6+N)\times SL(2,\R)$.
The same considerations as given above show that the stability
subgroup of the $O(6,6+N)$ factor is $O(1,5)\times O(1,5+N)$.  The
stability subgroup of the $SL(2,\R)$ factor, which would be $O(2)$ in
a standard Euclidean-torus reduction, will now instead be $O(1,1)$.
The reason for this is that the kinetic term for the axion $\chi$ in
(\ref{duald4extra}) will be reversed when the dualisation from
(\ref{d4extra}) is performed in the four-dimensional
Euclidean-signatured space.  Thus the coset for the scalar manifold in
$D=4$ will be
\be
\fft{O(6,N+6)}{O(1,5)\times O(1,N+5)}\, \times\,
\fft{SL(2,\R)}{O(1,1)}\ .
\ee

     In $D=3$, the global symmetry group is $O(8,N+8)$.  In this case,
upon setting all the axions to zero, the $(N+16)\times (N+16)$ matrix
$\cM$ is given by
\be
\cM= {\rm diag}\, (e^{\vec c_\sA\cdot\vec\phi}\, \delta^\sB_\sA\, ,
    \, \delta^\J_\I \, , \, 
          e^{-\vec c_\sA\cdot\vec\phi}\, \delta^\sA_\sB)\ ,\label{3cmu}
\ee
where $2\le A\le 9$ (see section \ref{ssec:d3coset}).  The field
redefinition (\ref{complex}) implies that there will now be four terms
in (\ref{3cmu}) that undergo sign reversals, namely those
corresponding to $A=2$ and $A=9$.  Thus, setting the dilatons to zero,
we obtain a fiducial matrix of the form (\ref{fiducial}) with $m=n=2$.
Consequently, the coset space describing the theory in $D=3$ obtained
by timelike reduction has the form
\be
\fft{O(8,N+8)}{O(2,6)\times O(2,N+6)}\ .
\ee

     The coset structures for Lorentzian torus reductions that we
have derived in this section are all in agreement with those given in
\cite{hj}.

\section{K3 compactifications of M-theory\label{sec:k3}}

Now let us consider, by contrast, the K3 reduction of $D=11$
supergravity.  We aim to show in some detail the relations between
this theory and the $T^3$ reduction of the $E_8\times E_8$ heterotic
theory as discussed above, which are conjectured to be equivalent
under duality \cite{ht,wit1}. The numbers of zero modes in the two
theories may be compared straightforwardly by counting the harmonic
forms corresponding to the various $D=7$ spin sectors in each theory.

Reducing $D=11$ supergravity on K3, one obtains a $D=7$ metric and 58
scalars from the reduction of the $D=11$ metric, plus a $D=7$
three-form antisymmetric tensor gauge potential plus 22 one-form gauge
potentials from the reduction of the $D=11$ three-form gauge
potential. The 58 scalars arise from the 57 shape-determining plus one
volume-setting moduli of the internal K3 manifold, while the 22
one-forms arise from the 22 harmonic two-forms occurring on K3.

In the supergravity sector of the heterotic theory compactified on
$T^3$, one has for comparison: a $D=7$ metric plus a triplet of
dilatonic scalars, together with a triplet of one-form Kaluza-Klein
gauge potentials and a triplet of axionic scalars, all descending from
the $D=10$ metric; one additional dilatonic scalar descending from the
$D=10$ dilaton; and also a $D=7$ two-form gauge potential, three
one-form gauge potentials and three axionic scalars, all descending
from the $D=10$ two-form gauge potential. The Yang-Mills sector of the
heterotic theory contributes a number of $D=7$ zero-modes as well. We
shall assume for this accounting that the Yang-Mills gauge symmetry is
``fully Higgsed,'' \ie that the gauge symmetry is maximally broken by
giving vacuum expectation values to the various $E_8\times E_8$
adjoint representation scalar fields.  This breaks the $E_8\times E_8$
group down to its Cartan subgroup, $\left(U(1)\right)^{16}$.  Thus we
get 16 further vector potentials and $16\times 3 = 48$ axionic scalars
from the Yang-Mills sector.  Comparing with the K3 reduction of
M-theory as described above, we see that we have a total of 22 vector
potentials in each case, and 58 scalars in each case.  From the K3
reduction we have a 3-form potential in $D=7$, while from the $T^3$
reduction of the heterotic theory we have a 2-form potential instead.
This indicates that one needs to perform a dualisation of one of the
two seven-dimensional theories in order to make contact with the
other.  In particular, this indicates that the relation between the
two involves an interchange between strong and weak coupling regimes.

To make things more precise, let us begin by looking in detail at the
$T^3$ reduction of the heterotic theory. In order to make contact with
the K3 reduction of $D=11$ supergravity, we will make a dualisation of
the 2-form potential $A_\2$ arising in the $T^3$ reduction of the
heterotic theory.  To do this, we need to know the Bianchi identity
for the field strength $F_\3$. From the results given in Appendix
\ref{app:hetred}, we find this to be $dF_\3 +F_{\2\a}\wedge \cF_\2^\a
+\ft12 G_\2^I\wedge G_\2^I=0$.  To dualise $A_\2$, we introduce a
3-form $A_\3$ as a Lagrange multiplier, adding the term $A_\3\wedge
(dF_\3 +F_{\2\a}\wedge \cF_\2^\a +\ft12 G_\2^I\wedge G_\2^I)$ to the
undualised Lagrangian.  Treating $F_\3$ now as an auxiliary field, we
can solve its algebraic equation of motion, giving $e^{\vec
  a_1\cdot\vec\phi}\, {*F_\3} = dA_\3 \equiv F_\4$.  Substituting this
back into the Lagrangian, we obtain the dualised version
\bea
e^{-1}\, {\cal L}_{7} &=& R  -\ft12
(\del\vec\phi)^2 -\ft1{48} e^{-\vec a_1\cdot\vec\phi}\, (F_\4)^2 -
\ft14 \sum_\a e^{\vec a_{1\a}\cdot\vec\phi}\, (F_{\2\a})^2 \nn\\
&&
-\ft12 \sum_{\a<\b} e^{\vec a_{1\a\b}\cdot\vec\phi}\, 
(F_{\1\a\b})^2 -\ft12 \sum_I e^{\vec c\cdot\vec\phi}\, 
(G_\2^I)^2 -\ft12 \sum_{\a, I} e^{\vec c_{\a}\cdot\vec \phi} \, 
 (G_{\1\a}^I)^2 \nn\\
&& -\ft14 \sum_{\a} e^{\vec b_\a\cdot\vec\phi}\, 
(\cF_\2^\a)^2 -\ft12 \sum_{\a<\b} e^{\vec b_{\a\b}\cdot\vec\phi}\, 
(\cF^\a_{\1\b})^2 \label{d7dualhet}\\
&&+ {*\Big(}A_\3\wedge (d{A'}_{\1\a}\wedge d\hA_\1^\a + 
 \ft12 d{{B^I}'}_\1 \wedge d{{B^I}'}_\1)\Big)\ ,\nn
\eea
 where $F_\4=dA_\3$, and the other field strengths are given in
(\ref{newfields}). 

To make comparison with the K3 reduction of $D=11$ supergravity, we
first need to discuss the nature of the K3 manifold.  (A detailed
account of its properties may be found in \cite{asp}.)  The required
K3 metric is Ricci flat and K\"ahler.  Although an existence proof for
Ricci-flat metrics on K3 has been given long ago \cite{yau}, the
explicit form of such metrics is still unknown, owing to the
complexity of the Einstein equation.  It is, however, possible to give
an approximate construction of the Ricci-flat metrics (see, for example,
\cite{gp,page}).  A
detailed discussion of this ``physical'' picture was given in
\cite{page}.  One can construct K3 by beginning with the 4-torus
$T^4$, defined by identifying coordinates $y^i\sim y^i + 2\pi$ in
$\R^4$.  Next, we make the identification $y^i\sim -y^i$.  This
identification has 16 fixed points, located at $y^i=\pi\, n_i$, where
$n_i$ are any integers.  One then cuts out a small 4-ball around each
of the 16 fixed points.  Had we not performed the identifications, the
boundaries of the 4-balls would each have been a 3-sphere.  Because of
the identification, the boundaries are instead copies of $RP^3$, the
real projective plane.  (This is $S^3$ with an antipodal
identification.)  One now patches up the manifold, by ``plugging in''
an appropriate space into each of the 4-balls.  What is needed is a
smooth 4-space with curvature localised to a small region, and which
then opens out into an asymptotic region that approaches flat
Euclidean 4-space but with an antipodal identification so that its
boundary is again $RP^3$.  Such a space is known; it is the
Eguchi-Hanson instanton \cite{eh}, which indeed approaches Euclidean
space factored by $Z_2$ at infinity \cite{bgpp}. By taking the
``size'' of the instanton to be sufficiently small, one achieves an
almost-smooth join between the torus and the Eguchi-Hanson instanton,
which splays out like a champagne cork and plugs into the excised
4-ball in the 4-torus.  Inserting a total of 16 such corks, \ie one
for each excised 4-ball, one achieves an approximation to a Ricci-flat
K3 manifold that becomes arbitrarily precise as the scale sizes of the
Eguchi-Hanson instantons are taken to zero \cite{page}.

     Using the above construction, it is possible to give a fairly
explicit construction of the K3 compactification of $D=11$
supergravity.  In particular the harmonic 2-forms on K3, in terms of
which all of the non-trivial zero-modes are described, can be seen to
fall into two different categories.  First of all, there are those
that can be viewed as being harmonic 2-forms on the 4-torus.  In fact,
one can easily see that the set of harmonic forms on the 4-torus that
will be present in the K3 itself will be the subset that survives the
antipodal identification $y^i\sim -y^i$.  Thus the six harmonic
2-forms $dy^i\wedge dy^j$ survive, whilst the four harmonic 1-forms
$dy^i$ are projected out.  Thus we may define three self-dual 2-forms
$J^\a_+$ and three anti-self-dual 2-forms $J^\a_-$ on $T^4$ (with the
triplet index $\a$ running over the values $2,3,4$ in order to match
with the notation arising in the $T^3$ compactification of the
heterotic string):
\bea
J^2_\pm &=& dy^1\wedge dy^4 \pm dy^2 \wedge dy^3\ ,\nn\\
J^3_\pm &=& dy^2\wedge dy^4 \pm dy^3 \wedge dy^1\ ,\label{sixj}\\
J^4_\pm &=& dy^3\wedge dy^4 \pm dy^1 \wedge dy^2\ ,\nn
\eea

The second type of harmonic 2-forms are those associated with the
Eguchi-Hanson instantons.  There is one of these for each of the
sixteen instantons, corresponding to the fact that the Eguchi-Hanson
solution has one normalisable harmonic 2-form.  We shall represent
these 2-forms by the symbols $\omega_\2^I$.  Each of these 2-forms is
strongly localised within a small region of the Eguchi-Hanson space
itself.  Concretely, the Eguchi-Hanson metric is given by \cite{eh}
\be
ds^2 = \Big(1-\fft{a^4}{r^4} \Big)^{-1}\, dr^2 + \ft14 r^2\,
\Big(1-\fft{a^4}{r^4}\Big)\, (d\psi + \cos\theta\, d\varphi)^2 
+ \ft14 r^2\, (d\theta^2 + \sin^2\theta\, d\varphi^2)\ .
\ee
 The radial coordinate has the range $a\le r <\infty$, with the bolt
occurring at $r=a$.  The coordinates $\theta$ and $\varphi$ are
angles on $S^2$.  The level surfaces $r = {\rm constant}$ have the
topology of $RP^3$, since the fibre coordinate $\psi$ has period $2\pi$
rather than the $4\pi$ period that would occur on $S^3$ \cite{bgpp}.  In
the natural orthonormal frame $e^0= (1-a^4/r^4)^{-1/2}\, dt$, $e^1=
\ft12 r\, d\theta$, $e^2 = \ft12 r\, \sin\theta\, d\varphi$, $e^3=\ft12 r
(1-a^4/r^4)^{1/2}\, (d\psi+\cos\theta\, d\varphi)$, it is easy to see
that the anti-self-dual 2-form
\be
\omega_\2 = \fft1{r^4} (e^0\wedge e^3 - e^1\wedge e^2)
\ee
 is closed.  This is the normalisable anti-self-dual harmonic 2-form on
Eguchi-Hanson.   

   In total, we then have 22 harmonic 2-forms; six from $T^4$ plus 16
from the corks.  These divide into 3 self-dual harmonic 2-forms from
$T^4$, plus $19 = 3+16$ anti-self-dual harmonic 2-forms, coming from
$T^4$ and from the 16 Eguchi-Hanson metrics.  As we shall see, the
ways in which the 2-forms from $T^4$ and from the Eguchi-Hansons
contribute in the dimensional reduction procedure will be slightly
different.

   Upon performing the K3 reduction, the $D=11$ fields give rise
to the following $D=7$ fields:
\bea
g_{\sst{MN}} &\longrightarrow & g_{\mu\nu}\ , \qquad \cA^i_{\0 j}\ ,
  \qquad \cA^I_{\0 \a}\ ,\qquad \vec\phi\ ,\nn\\
A_\3   &\longrightarrow & A_\3\ ,\qquad A_{\1 ij}\ ,\qquad 
               A_{\1 I}\ .\label{d7k3fields}
\eea
 Here, as usual, the $i,j,\ldots$ indices range over the four internal
coordinates $y^i$.  There are four dilatons $\vec\phi$, arising from
the fact that in the construction of K3 that we are using here, there
are the usual four circles making up the 4-torus.  There are also six
axions $\cA^i_{\0 j}$ ($i<j$) that parameterise the angular
deformations of the 4-torus.  There are also $48=16\times 3$ further
scalars $\cA^I_{\0 \a}$, corresponding to the remaining parameters that
make up the total of $58=10+48$ moduli of K3 \cite{hp,hitch}.  These
can be understood in the picture of K3 that we are using as follows.
There are two parameters that characterise the orientation of each of
the Eguchi-Hanson instantons, and a further parameter characterising
its scale size.  This gives $16\times (2+1)=48$ parameters in total
that are associated with the Eguchi-Hanson corks.  The 22 vector
fields that we mentioned previously split as six vectors $A_{\1 ij}$
coming from the $T^4$ reduction of $A_3$, plus 16 vectors $A_{\1 I}$
coming from the harmonic expansion involving the 16 harmonic 2-forms
$\omega_\2^I$ localised in the 16 Eguchi-Hanson instantons.

     The dilaton couplings for each field can be obtained
straightforwardly, by examining the ansatz for the reduction of the
eleven-dimensional metric on $T^4$:
\be
ds_{11}^2 = e^{\fft13\vec g\cdot\vec\phi}\, ds_7^2 + \sum_{i=1}^4
e^{2\vec \gamma_i\cdot\vec\phi}\, (dz^i + \cA^i_{\0 j}\, dz^j )^2\ .
\label{d11d7red}
\ee
 The constant vectors $\vec g$ and $\vec \gamma_i$ can be found in
\cite{cjlp1,lpsol}, and are given by
\bea
&&\vec\gamma_1 =(-\ft23, 0,0,0) \ ,\qquad 
  \vec\gamma_2 =(\ft1{12},-\ft{\sqrt 7}{4}, 0,0)\ ,\nn\\
&&\vec\gamma_3 =(\ft1{12}, \ft1{4\sqrt7}, -\sqrt{\ft37}, 0)\ ,\qquad
  \vec\gamma_4 =(\ft1{12}, \ft1{4\sqrt7},
  \ft1{2\sqrt{21}},-\sqrt{\ft5{12}})\ ,\label{gammas}\\
&&\vec g= -\ft65 \sum_i\vec\gamma_i = (\ft12, \ft3{2\sqrt7},
\sqrt{\ft37}, \sqrt{\ft35})\ .\nn
\eea
 Note that the radius $R_i$ of the $i$'th circle on $T^4$, and the $T^4$
volume $V_4=\prod_i R_i$, are given by 
\be
R_i=e^{\vec\gamma_i\cdot \vec\phi}\ ,\qquad V_4 = e^{-\fft56\vec
  g\cdot\vec\phi} \ .\label{radvol}
\ee
 Thus the combination of dilatons $\varphi=\sqrt{5/8}\, 
\vec g\cdot\vec\phi$ is the breathing mode of $T^4$, and hence also of
K3.  It is also sometimes useful to present another form of the 
metric reduction
ansatz that is applicable to computations that do not involve the
``internal'' structure of the compactifying 4-manifold, but only depend
on the breathing mode.  This is given by
\be
ds_{11}^2 = V_4^{-2/5}\, ds_7^2 + V_4^{1/2}\, ds_4^2\ .\label{d11d7red2}
\ee

   Associating dilaton vectors with the various seven-dimensional
fields as follows,
\be
\matrix{ \cA^i_{\0 j} & \cA^I_{\0 \a} & A_\3 & A_{\1ij} & A_{\1 I} \cr
         \vec b_{ij} & \vec b_\alpha & \vec a & \vec a_{ij}& \vec d}
\ ,\label{k3dv}
\ee
 we find that they can be expressed in terms of $\vec g$ and $\vec
\gamma_i$ as
\bea
&&\vec b_{ij} = 2\vec\g_i -2\vec\g_j\ ,\qquad \vec b_2 = \vec \g_1 +
\vec \g_2 - \vec \g_3 - \vec \g_4\ ,\nn\\
&& \vec b_3 = \vec \g_2 + \vec \g_3 - \vec \g_1 - \vec \g_4\ ,\qquad
\vec b_4=\vec \g_3 + \vec \g_1 -\vec \g_2 -\vec \g_4\ ,\nn\\
&&\vec a=-\vec g\ ,\qquad \vec a_{ij} = -2\vec\g_i-2\vec\g_j 
-\ft13\vec g \ ,\qquad \vec d =\ft12\vec g\ .\label{k3abd}
\eea

We shall now show in detail how these dilaton couplings arise in the
K3 reduction of M-theory.  The dilaton vectors $\vec a$, $\vec a_{ij}$
and $\vec b_{ij}$, corresponding to the fields $A_\3$, $A_{\1ij}$ and
$\cA^i_{\0 j}$ can be understood straightforwardly, since they are a
subset of those one would obtain by dimensionally reducing M-theory on
$T^4$.  As we shall see, they are in fact nothing but an $SL(4,\R)$
truncation of maximal supergravity in $D=7$, which has an $SL(5, \R)$
global symmetry.  To see this, consider the ansatz for the reduction
of the 3-form potential
\be
A_\3(x,y) = A_\3(x) + \ft12 A_{\1ij}(x)\wedge dy^i\wedge dy^j + 
A_{\1 I}(x)
\wedge \omega_\2^I\ .\label{a3red}
\ee
 For now, it is only the first two terms here that concern us.  From
the metric ansatz (\ref{d11d7red}), we can see that the determinant of
the vielbein reduces according to $e\rightarrow e\, e^{\fft13\vec
g\cdot\vec\phi}$, and thus we have that
\bea
- \ft1{48}e\, F_\4^2 \longrightarrow -\ft1{48}e\,  e^{-\vec
  g\cdot\vec\phi}\, F_\4^2 -\ft1{4} \sum_{i<j} 
  e^{-(2\vec\g_i+2\vec\g_j+\fft13
  \vec g)\cdot\vec\phi}\, (F_{\1ij})^2 +\cdots
\eea
 where $\cdots$ represents the $F_{\2 I}$ terms that we shall discuss
presently, and in obtaining the exponential factors we have used the
appropriate inverse metric components in $D=7$ or $D=4$, as given in
(\ref{d11d7red}).  The exponents can indeed be seen to be $\vec
a\cdot\vec\phi$ and $\vec a_{ij}\cdot\phi$ in (\ref{k3abd}).  The
dilaton couplings for the axions $\cA^i_{\0j}$ coming from the torus
reduction of the metric follow from the standard Kaluza-Klein formulae
as given, for example, in \cite{lpsol}.  In fact we can also obtain
the result for these axions by a simple linearised calculation, and it
is useful to present this here because a similar argument will be used
below in discussing the more difficult case of the other 48 axions
coming from the K3 metric moduli. If a metric $\bar g_{ij}$ is
subjected to a transverse traceless perturbation $h_{ij}$, \ie $g_{ij}=\bar
g_{ij} + h_{ij}$ where $\bar g^{ij}h_{ij}=0$ and $\bar\nabla^i\,
h_{ij}=0$, then the perturbed Ricci
tensor will be of the form $R_{ij}=\bar R_{ij} + \ft12 \Delta_L
h_{ij}$, where the Lichnerowicz operator $\Delta_L$ is defined by
$\Delta_L\, h_{ij} = -\bar{\square} \, h_{ij} -2\bar R_{ikj\ell} \, 
h^{k\ell} + 2 \bar R_{(i}{}^k\, h_{j) k}$.  
The fluctuation $h_{ij}$ will therefore give rise to a
contribution of the form $\ft12 h^{ij}\, \square h_{ij}$ in the 
Einstein-Hilbert Lagrangian, 
where $h^{ij}=\bar g^{ik}\, \bar
g^{j\ell}\, h_{k\ell}$.  In the present context, we see from
(\ref{d11d7red}) that the metric fluctuation corresponding to the
axion $\cA^i_{\0 j}$ is given by $h_{ij} =
e^{2\vec\g_i\cdot\vec\phi}\, \cA^i_{\0 j}$ when $i<j$ (together with
$h_{ji}=h_{ij}$).  Thus we will get a contribution of the form $\ft12
e^{-2(\vec\gamma_i+\vec\gamma_j)\cdot\vec\phi} \,
e^{2\vec\g_i\cdot\vec\phi}\, \cA^i_{\0 j}\,
\square(e^{2\vec\g_i\cdot\vec\phi}\, \cA^i_{\0 j})\sim -\ft12
e^{2(\vec\g_i-\vec\g_j)\cdot\vec\phi}\, (\del\cA^i_{\0j})^2$.  This
result shows that indeed the dilaton vector describing the dilaton
coupling is the one given by $\vec b_{ij}$ in (\ref{k3abd}).

It is a little more involved to understand the dilaton vectors $\vec
d$ and $\vec b_\a$ describing the couplings of the 16 vector
potentials $A_{\1 I}$ and the 48 axions $\cA^I_{\0\a}$, since these
arise from the sixteen 2-form harmonics $\omega^I_\2$ that are
intrinsic to K3.  Let us begin by considering the vector fields
$A_{\1I}$, which are the easier of the two sets to analyse. From the
last term in (\ref{a3red}), we see that these give the  
$dA_{\1I}\,\wedge \omega^I_\2$ contributions to the eleven-dimensional
4-form $F_\4$.  In the same spirit as above, we can calculate the
dimensional reductions of these terms by making the necessary
contractions of indices using the appropriate metric components as given
by the Kaluza-Klein ansatz.  For these fields, since their internal
components involve $\omega_\2^I$, we should use the metric given by
(\ref{d11d7red2}).  Thus we find
$-\ft1{48}e\, F_\4^2 \rightarrow -\ft14 e\, V_4^{-2/5}\, V_4^{4/5}\,
V_4^{-1}\, (F_{\2I})^2 = -\ft14 e\, e^{\ft12\vec g\cdot\vec\phi}\,
(F_{\2I})^2$, which indeed agrees with the dilaton vector $\vec d$ given
in (\ref{k3abd}).

The determination of the dilaton vectors for the $\cA^I_{\0 \alpha}$
is more complicated.  One approach is to note that the
subgroup $GL(4,\R)\sim O(3,3)\times \R$ of the $O(3,19)\times \R$ global
symmetry group of the K3 reduction corresponds precisely to the unbroken
general coordinate symmetry on the $T^4$.  The antipodal
identification in the $T^4$ described above preserves this global
symmetry group.  All of the fields should therefore form representations
under this $GL(4,\R)$.  This can be seen in particular in the couplings
of the dilatonic scalars.  Specifically, the dilaton vectors should form
weight vectors under $GL(4,\R)=\R\times SL(4,\R)$.  Here the $\R$
factor is generated by the breathing mode.  Indeed, the dilaton
couplings of $F_\4$ and the $F_\2^\I$ depend only on the breathing
mode, and they are accordingly singlets under $SL(4,\R)$.  On the other
hand, the dilaton vectors $\vec a_{ij}$ of $F_{\2ij}$ are precisely the
weight vectors of the six-dimensional representation of $SL(4,\R)$,
after the subtraction of a universal constant vector associated with the
breathing mode.  The dilaton vectors $\vec b_{ij}$ for the scalars
$\cA^i{}_{\0j}$ form the positive roots of $SL(4,\R)$, with simple
roots $\vec b_{12}$, $\vec b_{23}$ and $\vec b_{34}$.  To see
that these axionic scalars, taken together with the dilatonic degrees of
freedom orthogonal to the breathing mode, realise the full $SL(4,\R)$,
it is necessary to include the negatives of the dilaton vectors, {\it
i.e.}\ $-\vec b_{ij}$, since the set of vectors $\pm \vec b_{ij}$ form
the complete root system of $SL(4,\R)$.  Note, however, that dilatonic
couplings with $-\vec b_{ij}$ do not occur in the Lagrangian.  This is a
reflection of the fact that the scalars parameterise the coset
$SL(4,\R)/O(4)$, and thus provide a {\it non-linear} realisation of
$SL(4,\R)$.  The negative roots are generated \cite{lpsweyl} by the
non-linear (Weyl group) transformation $\cA^i_{\0j} \rightarrow e^{-\vec
b_{ij}\cdot\vec\phi}\, \cA^i_{\0j} + \ldots$.  Thus for the scalar
sector, both the dilaton vectors and their negatives should be included
in discussing the global symmetry.

In the approximate description of K3 that we are using here, we may note
that each of the 16 fixed points under antipodal identification should
be ``patched'' with an Eguchi-Hanson instanton.  The 16 instantons are
equivalent, and so we can discuss just a single one of them as a
representative.  Each instanton contributes three metric zero modes,
described by three axions.  The insertion of the Eguchi-Hanson
instantons preserves the $SL(4,\R)$ symmetry of the original $T^4$,
since, as we have observed, its asymptotic limit is the same as the
antipodally-identified $T^4$.  Following the above discussion, the three
dilaton vectors and their negatives form a six-dimensional
representation of $SL(4,\R)$.  (The inclusion of the negatives of the
dilaton vectors is clearly necessary since there exists no triplet
representation of $SL(4,\R)$.)  We now find that the set $\pm \vec b_\a$
form a six-dimensional representation of $SL(4,\R)$, given the
chosen basis $\vec b_{ij}$ for the positive roots. 
This representation is not unique, however; another example is the
set $\vec a_{ij}$ discussed above.  However, the set $\pm \vec b_\a$
given in (\ref{k3abd}) forms the unique solution in which three vectors
together with their negatives comprise a six-dimensional representation. 
Thus we see that the forms of these dilaton vectors are dictated by the
global symmetry $SL(4,\R)$.

An alternative way to understand these dilaton vectors is to note that
the K3 metric has 3 self-dual and $3+16$ anti-self dual harmonic
2-forms.  In the approximate K3 construction, we see that for each
Eguchi-Hanson instanton there is one localised anti-self-dual harmonic
2-form
$\omega_\2$, and three covariantly-constant self-dual 2-forms
$J^\a_+$.  As shown in \cite{hp}, one can use these to build three
zero-mode deformations of the metric ({\it i.e.}\ Lichnerowicz
zero-modes) of the form $h_{ij} = J^a_{+ik}\, \omega^k{}_j$.  This
gives a total of $3\times 16=48$ metric zero-modes, which, together
with the 10 coming from the 4-torus (4 dilatons $\vec\phi$ plus 6
axions $\cA^i_{\0j}$), give the 58 metric zero-modes of K3.  In the
bulk $T^4/Z_2$ part of K3, we have a total of six 1-form field
strengths $\cF^i_{\1j}$ for six axions, forming three
pairs\footnote{Note that the indices 1, 2 and 3 on the $\cF_\1$ fields
are cyclic. In the Borel-type (Solvable Lie Algebra) gauge, the
Lagrangian is expressed naturally in terms of $\cF^i_{\1 j}$ with
$i<j$ and we have $\cF^i_{\1 j} = -\cF^j_{\1 i}\, e^{\vec b_{ji} \cdot
\vec \phi}$.}
$\cF_{\1\pm}^2 = \cF^1_{\1 4} \pm \cF^2_{\1 3}$, $\cF_{\1\pm}^3 =
\cF^2_{\1 4} \pm \cF^3_{\1 1} = \cF^2_{\1 4} \mp \cF^1_{\1 3}\,
e^{-\vec b_{13} \cdot \vec \phi}$, $\cF_{\1\pm}^4 = \cF^3_{\1 4} \pm
\cF^1_{\1 2}$. On the other hand, for each instanton there are just
three axions.  This difference is associated with the fact that, whereas
in the $T^4/Z_2$ bulk there are 3 self-dual and 3 anti-self-dual
constant harmonics associated with the above three pairs of axions,
there are in each instanton just three self-dual covariantly constant
2-forms.
    The nature of the dilatonic couplings for the three axions
associated with each instanton can be revealed by first studying the
detailed structure of the dilaton couplings for the bulk $T^4/Z_2$
pairs of field strengths $\cF^\a_{\1\pm}$.  Let us consider the
Lagrangian for the $\cF_{\1\pm}^2$ pair, which is given by
\bea
e^{-1}\, {\cal L} &=&-\ft12(\del\vec\phi)^2 
   -\ft14 e^{\vec b_{14} \cdot \vec\phi}\, (\cF^1_{\1
  4})^2 -\ft14 e^{\vec b_{23} \cdot \vec \phi}\, (\cF^2_{\1 3})^2
\nn\\ 
&=&-\ft12(\del\vec\phi)^2 
 -\ft14 e^{\ft12(\vec b_{14} + \vec b_{23}) \cdot \vec\phi}\, 
(\cF^2_{\1+}, \cF^2_{\1-})\pmatrix{c & s\cr s & c\cr}
\pmatrix{\cF^2_{\1+}\cr
                                     \cF^2_{\1-} \cr}\ ,\label{xxx1}
\eea
 where $c=\cosh\theta$ and $s=\sinh\theta$, with $\theta= \ft12(\vec
b_{14}-\vec b_{23})\cdot\vec \phi$.  Thus for $\cF^2_{\1+}$, the
dilaton couplings are naturally described by $\ft12(\vec b_{14}+ \vec
b_{23})$.  This is because we can set consistently set both $\cF^2_{\1
-}=0$ and $\theta=0$, and then $\cosh\theta$ is replaced by unity in the
dilaton coupling.  Indeed, the natural augmentation of (\ref{xxx1}) to
include additional axions $\psi_\I$ gives
\be
e^{-1}\,{\cal L} = -\ft12(\del\vec\phi)^2 
 -\ft14 e^{\ft12(\vec b_{14} + \vec b_{23}) 
              \cdot \vec\phi}\, 
\Big[ (\cF^2_{\1+}, \cF^2_{\1-})\pmatrix{c & s\cr s & c\cr}
\pmatrix{\cF^2_{\1+}\cr
                                     \cF^2_{\1-} \cr} + 
   \sum_\I (\del\psi_\I)^2\Big] \ .\label{on1}
\ee
 In other words, the original pair of axions not only have an overall
dilaton factor $e^{\ft12(\vec b_{14} + \vec b_{23}) \cdot \vec \phi}$
but also the $\pmatrix{c&s\cr s&c}$ matrix coupling, while all further
axions $\psi_i$ are ``unpaired,'' and have only the overall
$e^{\ft12(\vec b_{14} + \vec b_{23})\cdot\vec\phi}$ factor.  Thus we
can argue that one of the three axions associated with a given
Eguchi-Hanson instanton can naturally be grouped with the
$\cF^2_{\1\pm}$, with a Lagrangian contribution of the same form as
those of the $\psi_\I$ in (\ref{on1}).  This is because the three axions
can be approximately viewed as an internal self-dual truncation
(analogous to setting $\cF^2_{\1 -}$ to zero), since there are only
three self-dual constant harmonic 2-forms in the Eguchi-hanson metric.
This implies that the dilaton coupling for one of the three axions is
given by $\ft12(\vec b_{14} + \vec b_{23})=\vec \g_1 + \vec \g_2 - \vec
\g_3 -\vec \g_4$. The other two sets of $N=16$ axions are then associated
with the
$\cF^3_{\1\pm}$ and $\cF^4_{\1\pm}$ pairs, with dilaton couplings
given by $-\vec \g_1 + \vec\g_2 + \vec\g_3 -\vec \g_4$ and $\vec \g_1
-\vec \g_2 +\vec\g_3 - \vec\g_4$ respectively.

   Comparing with the $T^3$ reduction of the heterotic string that we
obtained previously, we find that the correspondence between the
fields in the two descriptions can be summarised in the following
Table:

\bigskip\bigskip
\centerline{
\begin{tabular}{|c|c|c|c|c|}\hline
   \multicolumn{2}{|c|}{M-theory on K3} & &
   \multicolumn{2}{|c|}{Heterotic string on $T^3$} \\ \hline
   $D=11$ & $D=7$ & Duality & $D=7$ & $D=10$ \\ \hline\hline
  & $A_\3$, \ \ \ $\vec a$ & $\longleftrightarrow$ 
  & $A_\2$, \ \ \ $\vec a_1$ & $A_\2$ 
   \\ \cline{2-4}
 & $A_{\1 14}$, \ \ \ $\vec a_{14}$ & $\longleftrightarrow$ &
          $A_{\1 2}$, \ \ \ $\vec a_{12}$ & \\
 & $A_{\1 24}$, \ \ \ $\vec a_{24}$ & $\longleftrightarrow$ & 
          $A_{\1 3}$, \ \ \ $\vec a_{13}$ & \\
 & $A_{\1 34}$, \ \ \ $\vec a_{34}$ & $\longleftrightarrow$ & 
          $A_{\1 4}$, \ \ \ $\vec a_{14}$ & \\ \cline{2-5}
$A_\3$ & $A_\1^{ I}$, \ \ \ $\vec d$ & $\longleftrightarrow$ &
  $B_\1^{ I}$, \ \ \ $\vec c$ & $B_\1^{ I}$ \\ \cline{2-5}
 & $A_{\1 12}$, \ \ \ $\vec a_{12}$ & $\longleftrightarrow$ & 
          $\cA^4_\1 $, \ \ \ $\vec b_4$ & \\
 & $A_{\1 13}$, \ \ \ $\vec a_{13}$ & $\longleftrightarrow$ & 
          $\cA_\1^3$, \ \ \ $\vec b_3$ & $G_{\mu\nu}$ \\
 & $A_{\1 23}$, \ \ \ $\vec a_{23}$ & $\longleftrightarrow$ & 
          $\cA_\1^2$, \ \ \ $\vec b_2$ & \\ \hline\hline\hline
 & $\cA_{\0 \a}^{I}$, \ \ \ $\vec b_\alpha$ &
           $\longleftrightarrow$
          & $B^{I}_{\0\a}$, \ \ \ $\vec c_\a$ & $B_\1^{I}$
         \\  \cline{2-5}
& $\cA^1_{\0 4}$, \ \ \ $\vec b_{14}$ &$\longleftrightarrow$ & 
       $A_{\0 34}$, \ \ \ $\vec a_{134}$ &  \\
& $\cA^2_{\0 4}$, \ \ \ $\vec b_{24}$ &$\longleftrightarrow$ & 
       $A_{\0 24}$, \ \ \ $\vec a_{124}$ & $A_\2$ \\
$G_{\mu\nu}$ & $\underline{\cA^3_{\0 4}}$, \ \ \ $\vec b_{34}$ 
&$\longleftrightarrow$ & 
       $\underline{A_{\0 23}}$, \ \ \ $\vec a_{123}$ & \\ \cline{2-5}
& $\underline{\cA^2_{\0 3}}$, \ \ \ $\vec b_{23}$ &$\longleftrightarrow$ & 
       $\underline{\cA^3_{\0 4}}$, \ \ \ $\vec b_{34}$ &  \\
& $\underline{\cA^1_{\0 2}}$, \ \ \ $\vec b_{12}$ &$\longleftrightarrow$ & 
       $\underline{\cA^2_{\0 3}}$, \ \ \ $\vec b_{23}$ &$G_{\mu\nu}$\\
& $\cA^1_{\0 3}$, \ \ \ $\vec b_{13}$ &$\longleftrightarrow$ & 
       $\cA^2_{\0 4}$, \ \ $\vec b_{24}$ & \\ \hline
\end{tabular}}
\bigskip

\centerline{Table 1: The correspondence between M-theory and the heterotic
  fields in $D=7$}
\bigskip\bigskip

In fact the fields $A_{\1ij}$ and $\cA^i_{\0 j}$ in the K3
compactification can be expressed in terms of the associated fields
of the heterotic theory in a more covariant fashion by making use of
the self-dual and anti-self-dual 2-forms $J^\a_\pm$ defined in
(\ref{sixj}):
\bea
A_{\1ij} &=& \ft12(J^\a_{+ij} + J^\a_{-ij})\, A_{\1\a} +
 \ft12(J^\a_{+ij} - J^\a_{-ij})\, \cA^\a_\1\ ,\nn\\
\cA^i_{\0 j} &=&  \ft14(J^\a_{+ij} + J^\a_{-ij})\,\ep_{\a\b\g}\, 
                       A_{\0\b\g} +
\ft14(J^\a_{+ij} - J^\a_{-ij})\,\ep_{\a\b\g}\, 
                       \cA^\b_{\0\g} \ .\label{k3hetrel}
\eea

The four dilatons in the heterotic and M-theory reductions are related
by an orthonormal transformation, $\vec\phi_{\rm H} = M\,
\vec\phi_{\rm M}$, where
\be
M=\pmatrix{
    \ft18 & \ft3{8\sqrt7} & \ft14\sqrt{\ft37} &
  \ft{\sqrt{15}}{4}\cr
\ft{11}{8\sqrt7} & -\ft{31}{56} & -\ft{31}{28\sqrt3} &
\ft14\sqrt{\ft5{21}} \cr
 -\ft5{2\sqrt{21}} & \ft{3}{14}\sqrt3 & -\ft{31}{42}
 &\ft16\sqrt{\ft57}\cr
 -\ft12\sqrt{\ft53} & \ft12\sqrt{\ft{15}{7}} &\ft16 \sqrt{\ft57}
 & \ft16}\ .\label{mmatrix}
\ee
 This matrix satisfies $M^5=\oneone$.  To understand the nature of
this transformation, we note that both M-theory on K3 and the
heterotic theory on $T^3$ have an $O(3,3)\sim SL(4,\R)$ global
symmetry as a subgroup of the full $O(3,19)$.  The simple roots of
$O(3,3)$ are given by the dilaton vectors of the underlined axionic
fields listed in Table 1.  Thus we have

\bigskip\bigskip
\centerline{
\begin{tabular}{ccccccccccc}
\multicolumn{5}{c}{M-theory on K3} & $\phantom{XXXXXXXX}$ &
\multicolumn{5}{c}{Heterotic string on $T^3$} \\
&&&&&&&&&& \\
$\vec b_{12}$ & & $\vec b_{23}$ & & $\vec b_{34}$ & &
$\vec b_{23}$ & & $\vec b_{34}$ & & $\vec a_{123}$ \\
o&---&o&---&o&   & o&---&o&---&o
\end{tabular}}
    
\bigskip
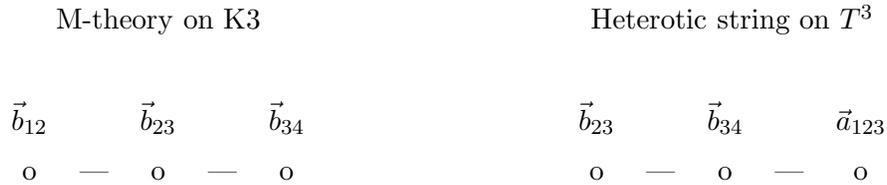
\centerline{Figure 1. Simple roots of the $O(3,3)$ subgroup}

\bigskip\bigskip

The $O(3,3)\sim SL(4,\R)$ group can be also viewed as a subgroup of
the $SL(5,\R)$ global symmetry group of maximal supergravity in $D=7$,
which has simple roots $\vec b_{12}$, $\vec b_{23}$, $\vec b_{34}$ and
$\vec a_{123}$.  Thus we see that M-theory on K3 and the heterotic
string on $T^3$ make two different truncations of $SL(5,\R)$. From the
maximal supergravity point of view, the two sets of simple roots of
$O(3,3)$ are related by the Weyl group of $SL(5,\R)$, which is $S_5$,
the permutation group of five objects.  Thus we would naturally expect
that $M^5=\oneone$.

The matrix $M$ also maps the dilaton vector $\vec a=- \vec g $ of the
4-form field strength $F_\4$ in the M-theory reduction to $-\vec a_1$
of the 3-form field strength $F_\3$ in the heterotic string reduction.
The minus sign is consistent with the fact that a dualisation of the
4-form field strength is necessary in order to make the identification
of the two theories.  Since the effective string coupling $\lambda_7$
of the heterotic string on $T^3$ is given by $\lambda_7=e^{-\ft58 \vec
a_1 \cdot \vec \phi}$, it follows from (\ref{radvol}) that the
seven-dimensional string coupling is \cite{wit1}
\be
\lambda_7 = V_4^{3/4} \ .\label{d7coupling}
\ee
The complete set of mappings for all the dilaton vectors can be seen
from Table 1.  Note that we have concentrated so far on establishing
the relation between the dilaton couplings in the two theories; the
detailed matching of the Kaluza-Klein modifications to the field
strengths requires a more detailed analysis.

    It is appropriate at this point to make a few remarks about the
nature of the Kaluza-Klein reduction procedure in the K3
compactification, and in particular to address the issue of the {\it
consistency} of the reduction.  In principle, the first step in any
Kaluza-Klein compactification is to perform a harmonic expansion of
all the higher-dimensional fields in terms of appropriate complete
sets of scalar and tensor harmonics on the internal space, thereby
arriving at a lower-dimensional theory with infinite towers of massive
fields, together with finite numbers of massless fields.  At this
stage the reduction is guaranteed to be consistent, since one has done
nothing more than a generalised Fourier expansion of the
higher-dimensional fields.

   In practice, one is usually interested in retaining only the finite
number of massless fields arising from the Kaluza-Klein reduction.  In
other words, one would ideally wish to be able to set the infinite
towers of massive fields to zero.  The question then arises as to
whether this is a consistent truncation of the lower-dimensional
theory.  In other words, is the setting to zero of the massive fields
consistent with their own equations of motion?  The dangers of
inconsistency all stem from the non-linear interaction terms in the
theory, which have the possibility in general of including ``source
terms'' for the massive fields, built purely from the massless fields
that are to be retained.  Thus if we denote the massless fields
generically by $\phi_{\sst L}$, and the massive ones by $\phi_{\sst M}$, 
a typical
inconsistency would be signalled by the occurrence in the Lagrangian
of non-linear interactions of the form $\phi_{\sst M}\, (\phi_{\sst L})^2$, 
leading
to equations of motion of the form
\be
\square \phi_{\sst M} + M^2 \, \phi_{\sst M} \sim (\phi_{\sst L})^2\ ,
\label{lm}
\ee
 which would not allow the massive fields to be set to zero.

        In the simplest cases, such as Kaluza-Klein reduction on a
torus, such dangerous terms cannot occur and so the truncation is
indeed consistent.  This follows from a simple group-theoretic
argument: if all the massless fields are singlets under some global
symmetry group, while all the massive fields are non-singlets, then
the massless fields cannot provide sources for the massive ones
\cite{dpconsist}.  In the toroidal case, the group in question is the
global $U(1)^n$ symmetry of the $n$-torus.  The massless fields are
all uncharged, and are hence singlets under the $U(1)$ factors, while the
massive fields are charged.  To put it another way, products of the
zero-mode harmonics on the torus (which are all constant) cannot
generate non-zero-mode harmonics.

    For the K3 reduction, the situation is less clear-cut.  It would
seem now to be quite conceivable that the product of zero-mode
harmonics could generate non-zero-mode ones, since even the zero-mode
harmonics are not now in general covariantly-constant.
Correspondingly, one might expect that the massive fields could now
have equations of motion of the form (\ref{lm}).  It was argued in
\cite{zilch,dfps} that, in K3 or Calabi-Yau compactifications of
supergravities, the source terms must in fact necessarily involve
derivative couplings, of the generic form
\be
\square \phi_{\sst M} + M^2 \, \phi_{\sst M} \sim 
(\del\phi_{\sst L})^2\ ,\label{lm2}
\ee
 and that there accordingly exists a regime of excitations of the massive
fields where the energies are small compared to the Kaluza-Klein
mass scale $M$, with the consequence that the inconsistencies in such
reductions can then be neglected. One can expand (\ref{lm2}) in such
cases as
\be
\phi_{\sst M} \sim M^{-2}\, \Big(1+ M^{-2}\, \square \Big)^{-1}\,
(\del\phi_{\sst L})^2 \sim M^{-2}\, (\del\phi_{\sst L})^2 +
\cdots\ ,\label{approx1}
\ee
 Thus, at low energies the $\square \phi_{\sst M}$ term in the massive
field equation can be dropped, and the massive field $\phi_{\sst
M}$ can effectively be ``integrated out,'' by substituting the solution
$\phi_{\sst M} \sim M^{-2}\, (\del\phi_{\sst L})^2$ into the lower-dimensional
Lagrangian.  A related approach is to substitute the Kaluza-Klein
reduction ansatz for the massless fields into the higher-dimensional
Lagrangian, and then to integrate over the internal compactifying
manifold.

     Such an approximate discussion is applicable to 
situations where one is seeking to extract an effective low-energy
``phenomenological'' theory from the string compactification, where
the mass scale $M$ of the Kaluza-Klein massive modes is very large
compared with the energies of interest.  However, it is not clear that
this applicability extends to the regime of interest for non-perturbative
duality symmetries.  In particular, the conjectured duality between the
heterotic string compactified on $T^3$ and M-theory compactified on K3
involves an inverse relation between the scale sizes of the $T^3$ and
K3.  Thus, to make any meaningful statements it is necessary to consider
M-theory compactified on a {\it large} K3, where the Kaluza-Klein mass
scale $M$ tends to zero, and here the neglect of kinetic terms for
massive fields such as that in (\ref{lm2}) becomes less and less
innocent.  Indeed, as $M$ tends to zero it is presumably more
appropriate to expand (\ref{lm2}) using not (\ref{approx1}), but rather
\be
\phi_{\sst M} \sim \Big(1+ M^{2}\, \square^{-1}\Big)^{-1}\, \square^{-1}\,
(\del\phi_{\sst L})^2 \sim  \square^{-1}\, (\del\phi_{\sst L})^2 
+\cdots\ .\label{approx2}
\ee
 Thus, rather than having the situation sketched in (\ref{approx1}) where
the effect of ``integrating out'' the massive modes is to modify the
effective low-energy action by higher-order derivative couplings that
are damped by inverse powers of the Kaluza-Klein mass scale, the
effect now in the small-$M$ regime is to obtain non-derivative
modifications with no damping.

    It is therefore important in the context of M-theory/heterotic
duality to try to establish whether or not the truncation to the
massless sector is a consistent one.  It is often asserted that the
modulus space for Ricci-flat metrics on K3 is the coset space
$O(3,19)/(O(3)\times O(19))$, and that this endows the
seven-dimensional theory following from the K3 reduction of M-theory
with a scalar manifold having this same coset structure.  Indeed, it
appears to be the case that if one substitutes the Kaluza-Klein ansatz
for the zero-mode sector into the $D=11$ Lagrangian, and then integrates
over K3, then the resulting seven-dimensional Lagrangian will have a
scalar sector described by this $O(3,19)/(O(3)\times O(19))$ coset.
However, as we have discussed above, it is a much more exacting and
stringent question as to whether instead the substitution of the
zero-mode ansatz into the eleven-dimensional field equations will be
consistent with these fields' own equations of motion.  Furthermore,
although the effects of integrating out massive fields in the
low-energy approximation (\ref{approx1}) would not upset the coset
structure of the Lagrangian for the lower-dimensional scalar fields, it
is not so clear that this sigma-model structure would survive unscathed
in the small-$M$ regime described by (\ref{approx2}).  As far as we are
aware, this is a question that has not been addressed in the literature,
and there appears to be no {\it a priori} argument that guarantees the
consistency of the truncation.  Although this could be argued to be a
negligible problem in the context of low-energy phenomenology, it
would seem to be a more significant one in the context of
M-theory/heterotic duality, and it is deserving of further study.
Indeed, one might argue that the consistency issue could provide
a non-trivial test of the validity of the
conjectured duality between M-theory and the heterotic string:
Since the truncation of the $T^3$ compactification of the heterotic
string to its massless sector is consistent, then the consistency of
the truncation to the massless sector of the K3 compactification of M-theory
would be a necessary consequence of the duality between the two theories. 

\section{Charge lattice relations\label{sec:chargelattice}}

In this section, we shall consider in detail the relation between the
lattices of electric and magnetic charges that are allowed by the
Dirac quantisation conditions in the K3 reduction of M-theory and the
$T^3$ reduction of the heterotic string. It has been shown in Refs
\cite{schwarzp,alwis,dirac} that the minimum charges of M-branes can
be fixed by invoking duality relations between M-theory, type IIA and
type IIB string theories, together with the existence of certain
``scale-setting'' $p$-brane species \cite{dirac}. In the case of
M-theory, the charge units can also be fixed by consideration of the
topological ${\cal L}_{\sst{FFA}}$ term in the Lagrangian
\cite{dlm}. We shall now show that the M-brane charges can also be
fixed by consideration of the conjectured duality relation between
M-theory compactified on K3 and heterotic string theory compactified
on $T^3$, and we shall show that the results are consistent with the
previous ones. We shall use the relations between charges under
Kaluza-Klein dimensional reduction given in Ref./ \cite{dirac}.  To
begin with, let us consider the heterotic string in ten dimensions.
Making the gravitational constant $\kappa_H$ and the string tension
$\a'$ explicit, we may write the low-energy effective action as
\be
e^{-1}\, {\cal L}_{10} = \kappa_H^{-2}\, \Big(
       R  - \ft12 (\del\phi_1)^2 
 -\ft12 e^{\phi_1}\, (F_\3)^2 - \ft12 e^{\fft12 \phi_1}\, 
 {\rm tr}\, (G_\2)^2 \Big) \ ,\label{d10het2}
\ee
 where the $E_8\times E_8$ gauge fields $G_\2$ are written in terms of
gauge potentials as
\be
G_\2=dB_\1 + \fft1{\sqrt{\a'}}\, {[} B_\1, B_\1 {]}\ .
\ee

     We shall first consider the subset of the theory corresponding to
the $O(3,3)$-invariant subsector of the complete $D=7$ reduction; in
other words, in the heterotic picture the Yang-Mills fields are not
yet to be included.  Equivalently, in the M-theory picture, the fields
in $D=7$ associated with the ``Eguchi-Hanson harmonics'' on K3 are not
yet to be included.  Let us assume that the volume of K3 is given by
$V_4=L_1^4$, whilst the volume of the $T^3$ in the heterotic picture
is $V_3=L_2^3$.  The duality of the two theories implies that
\be
\kappa_7^2 = \fft{\kappa_{11}^2}{L_1^4} = \fft{\kappa_H^2}{L_2^3} \ .
\label{kappas}
\ee
 In $D=11$, the most general solution for the minimum M-brane charges,
consistent with the Dirac quantisation condition, is given by
\be
Q_{e\4} = (2\pi)^\a\, \kappa_{11}^{4/3}\ ,\qquad
Q_{m\4} = (2\pi)^{1-\a}\, \kappa_{11}^{2/3}\ ,\label{mbranecharge}
\ee
 where the $e$ and $m$ subscripts indicate electric and magnetic
charges, and the $\4$ subscript indicates that they are carried by the
4-form field strength.  The constant $\a$ is as yet arbitrary.  

      Upon reduction to $D=7$ on K3, in the $O(3,3)$ subset the 4-form
reduces to $F_\4$ and $F_{\2ij}$.  It follows that their charges are
given by 
\be
Q_{m\4} =  (2\pi)^{1-\a}\, \kappa_{11}^{2/3}\ ,\qquad
Q_{m\2ij} = (2\pi)^{1-\a}\, \kappa_{11}^{2/3}\, L_1^{-2} \ .
\label{d7k3charge}
\ee
 (We need only list either electric or magnetic charges, since these are
related by the Dirac quantisation condition.)

     In the heterotic picture, on the other hand, the $O(3,3)$ subset of
fields is obtained from the $T^3$ reduction of the pure $N=1$ supergravity
multiplet.  The most general solutions for the $D=10$ string and 5-brane
charges are given by
\be
Q_{e\3} =  (2\pi)^{\b}\, \kappa_{H}^{3/2}\ ,\qquad
Q_{m\3} = (2\pi)^{1-\b}\, \kappa_{H}^{1/2}\ .\label{hetcharges}
\ee
 Upon making a $T^3$ reduction, the 3-form gives rise to $F_\3$ and
$F_{\2\a}$, together with 1-form field strengths for axions, which we
shall not consider here.  In addition, there are three Kaluza-Klein
2-forms
$\cF^\a_\2$.  The charges of these fields are given by
\be
Q_{e\3} =  (2\pi)^{\b}\, \kappa_{H}^{3/2}\, L_2^{-3}\ ,\qquad
Q_{m\2\a} =  (2\pi)^{1-\b}\, \kappa_{H}^{1/2}\, L_2^{-1}\ ,\qquad
Q_{m\2\a}^{\rm KK} = L_2\ .\label{d7charges2}
\ee
 In each case, the corresponding magnetic or electric dual charges
are related by the $D=7$ Dirac quantisation conditions. 
Note that the magnetic Kaluza-Klein charge is associated with a NUT
charge, and hence its charge unit is determined by topological
considerations.  

     The duality between M-theory reduced on K3 and the heterotic
theory reduced on $T^3$ implies that the charges carried by the
various fields in the two pictures should be equated, in accordance
with the equivalences of the corresponding fields as given in Table 1.
This gives rise to three independent equations:
\be
(2\pi)^{\b}\, \kappa_{H}^{3/2}\, L_2^{-3} =
(2\pi)^{1-\a}\, \kappa_{11}^{2/3}\ ,\qquad
(2\pi)^{1-\b}\, \kappa_{H}^{1/2}\, L_2^{-1}= L_2=
(2\pi)^{1-\a}\, \kappa_{11}^{2/3}\, L_1^{-2} \ .\label{chargerels}
\ee
 Together with the relation (\ref{kappas}) between the gravitational 
constants, we find that
\bea
\kappa_{11}^2 = \ft1{2\pi}\, L_1^6\, L_2^3 \ ,&&
\kappa_{H}^2 = \ft1{2\pi}\, L_1^2\, L_2^6\ , \nn\\
\a= \ft23\ , && (2\pi)^{4\b} = (2\pi)^3\, L_1^2\, L_2^{-2}\ .\label{parsol}
\eea
 From these, we see that the M-brane charges in $D=11$ are completely
determined, and must be given by
\be
Q_{e\4} = n\, (2\pi\kappa_{11}^2)^{2/3}\ ,\qquad
Q_{m\4} = m\, (2\pi\kappa_{11}^2)^{1/3}\ ,\label{mbraneint}
\ee
 where $n$ and $m$ are integers.  This conclusion is the same as that
obtained in \cite{dirac}, where T-duality between the type IIA and
type IIB theories was invoked.

   We now turn to a consideration of the Yang-Mills sector.  The
$E_8\times E_8$ gauge potentials may be expanded in terms of the
Cartan-subalgebra generators $H_I$ and the non-zero root generators
$E_a$ as $B_\1 = B_\1^I\, H_I + B_\1^a\, E_a$.  In particular we see
that with respect to the Cartan subalgebra gauge potentials, the
coupling of the non-zero root potentials takes the form $dB_\1^a\, E_a
+ (\a')^{-1/2}\, B_\1^I\wedge B_\1^a {[} H_I, E_a{]} = (dB_\1^a +
(\a')^{-1/2}\, \a^{I a}\, B_\1^I\wedge B_\1^a)\, E_a$.  Thus the
fields $B_\1^a$ interact with the $U(1)^{16}$ Cartan subalgebra
potentials {\it via} minimal coupling, with the 16 electric charges
given in terms of the components of the root vectors $\vec \a_a$.  The
basic units of electric Yang-Mills charge in the ten-dimensional
heterotic string are therefore given in terms of the 16 simple-root
vectors $\vec\a_i$ of $E_8\times E_8$, since all the other Yang-Mills
charges are expressible in terms of linear combinations of these with
integer coefficients.  To make this more precise, we choose to define
the magnetic charges by the integrals
\be
Q_m^I \equiv \int G_\2^I\ .\label{magdef}
\ee
 It follows from the form of the covariant derivative  $DB_\1^i=dB_\1^i +
(\a')^{-1/2}\, \a^{I i}\, B_\1^I\wedge B_\1^i$ that the Dirac
quantisation conditions will require the magnetic charges to lie
on the reciprocal lattice
\be
\vec Q_m = m_i\, \sqrt{\a'}\, \vec\mu^i\ ,\label{mhetcharge}
\ee
 where $m_i$ are integers and $\vec\mu^i$ are the fundamental weight
vectors, defined by $\vec\a_i\cdot \vec\mu^j = \delta_i^j$.

For comparison, we now consider the charge lattice for the Yang-Mills
sector coming from the reduction of M-theory on K3.  In particular, we
consider the charges under the 16 abelian 2-form fields arising from
the 16 anti-self-dual harmonic 2-forms on K3 which, in the approximate
discussion of section \ref{sec:k3}, were associated with the 16
Eguchi-Hanson instantons.  These are the fields that are conjectured
to be related by duality to the Cartan subalgebra of the $E_8\times
E_8$ symmetry of the heterotic string.

We may derive the charge lattice for these 16 fields in the M-theory
picture by using an abstract description of the cohomology of K3.
Specifically, one may introduce a set of sixteen 2-forms
$\sigma^i_{\2}$ that are ``dual'' to the sixteen anti-self-dual
harmonics $\td\omega_{\2 i}$, in the sense that that
\be
\int_{K3} \sigma^i_{\2}\wedge \td\omega_{\2 j} = \delta^i_j\ .\label{sigmas}
\ee
 The 2-forms $\sigma^i_{\2}$ are normalised so that their integrals
over the relevant 16 2-cycles $\Sigma_i$ in K3 are given by
\be
L_1^2\, \int_{\Sigma_i} \sigma^j_{\2} = \delta^j_i\ .\label{cycles}
\ee
 (We are taking the harmonic 2-forms $\td\omega_{\2i}$ to have the
dimensions of (Length)$^2$, and the dual 2-forms $\sigma_\2^i$ to have
dimensions (Length)$^{-2}$.  We use the dimensionful parameter
$L_1=V_4^{1/4}$ that we introduced earlier, where $V_4$ is the volume
of K3, in order to balance the dimensions.)  Then, one has the result
that \cite{asp}
\be
\int_{K3} \td\omega_i\wedge\td\omega_j = L_1^4\, M_{ij}\ ,\label{cartanm}
\ee
 where $M_{ij}$ is the Cartan matrix of $E_8\times E_8$.  This is given
in terms of the simple root vectors $\vec \a_i$ by $M_{ij}=\vec
\a_i\cdot\vec \a_j$.  From the above equations, we easily see that if
we expand $\td\omega_{\2 i}$ in terms of $\sigma^i_{\2}$, we have
$\td\omega_{\2 i} = M_{ji}\, \sigma^j_\2$, and hence that
\be
\int_{\Sigma_i} \td\omega_{\2 j} = L_1^2\, M_{ij}\ .\label{norm1}
\ee

     The 2-forms $\td\omega_{\2 i}$ do not have the appropriate
normalisation for giving canonical diagonal kinetic terms for the
associated spacetime 2-form field strengths in the Kaluza-Klein
reduction of $F_\4$.  To obtain the proper kinetic terms, we should now
define new linear combinations $\omega_\2^\I$ by
\be
\omega_\2^\I = \mu^{\I i}\, \td\omega_{\2 i}
\ee
 where $\mu^{\I i}$ denotes the set of 16 components of the fundamental
weight vectors $\vec\mu^i$ that we introduced previously.  The 2-forms
$\omega_\2^\I$ therefore satisfy the relations
\be
\int_{K3} \omega_\2^\I\wedge \omega_\2^\J = L_1^4\, \delta^{\I\J}\ ,\qquad
\int_{\Sigma_i} \omega_\2^\I = L_1^2\, \a^\I_i\ .\label{omegaints}
\ee
 Performing the Kaluza-Klein reduction
\be
F_\4 = F_\2^\I\, \omega_\2^\I\ ,
\ee
 we see from (\ref{omegaints}) that the normalisation of the
$\omega_\2^\I$ implies that the kinetic terms for the spacetime
2-forms $F_\2^\I$ will be canonical.  The magnetic charge $Q_{m\4}$ in
$D=11$ for a 5-brane wrapped around the 2-cycle $\Sigma_i$ in K3 will
therefore be given by
\be
Q_{m\4} = \int F_\4 =  Q_m^\I\, \int_{\Sigma_i} \omega_\2^\I 
 =  L_1^2 \, Q_m^\I\, \a^\I_i\ ,\label{e8e8}
\ee
 where $Q_m^\I = \int F_\2^\I$ is the magnetic charge of the resulting
3-brane in $D=7$.  (Recall that $L_1$ is the length scale-factor for
K3 with volume $V_4=L_1^4$.)  The 5-brane charges in $D=11$ have
already been determined in (\ref{mbraneint}), however.  Thus it
follows that consistency\footnote{It is worth noting that there is an
alternative basis for the 16 Eguchi-Hanson type anti-self-dual
harmonic 2-forms $\td\omega_i$, in which we have $\int_{\Sigma_i}
\td\omega_j =L_1^2\, \delta_{ij}$ and (\ref{cartanm}), rather than
(\ref{norm1}) and (\ref{cartanm}).  In this alternative basis,
(\ref{mk3charges}) is replaced by $\vec Q_m = n_i\, \vec a_i\, Q_0$,
where $n_i$ are integers.  This is entirely equivalent to
(\ref{mk3charges}), since the integers $n_i$ are related to the
integers $m_i$ in (\ref{mk3charges}) by $n_i = (M^{-1})^{ij}\, m_j$,
and both $M_{ij}$, the Cartan matrix of $E_8\times E_8$, and its
inverse $(M^{-1})^{ij}$, have entirely integer-valued components.}
with (\ref{e8e8}) requires that the magnetic 3-brane charges in $D=7$
lie on the lattice
\be
\vec Q_m = m_i\, \vec \mu^i\, Q_0 ,\label{mk3charges}
\ee
 where $\vec Q_m = (Q_m^1, Q_m^2, \ldots, Q_m^{16})$ and
\be
Q_0 = (2\pi\, \kappa_{11}^2)^{1/3}\, \Big( \fft{\kappa_H^2}{2\pi\,
  \kappa_{11}^4} \Big)^{1/5}\ .
\ee
 Note that the lattice of magnetic charges (\ref{mk3charges}), derived
from the K3 compactification of M-theory, is of the same form as the
lattice (\ref{mhetcharge}) that we obtained in the $T^3$ compactification
of the heterotic string.  A precise identification then requires that $\a'
= Q_0^2$.

\section{$p$-brane orbits with Yang-Mills charges}

     In the previous section, we looked at the duality relation
between the heterotic string compactified on $T^3$ and M-theory
compactified on K3.  As well as being a relation that holds in the
low-energy effective field theories, this should also be seen at the
level of BPS states in the two full theories. For example, the
M5-brane wrapped around the K3 manifold is dual to a vertical
reduction of the NS-NS string of the heterotic theory, while the
vertical reduction of the M2-brane is dual to the NS-NS 5-brane
wrapped around the volume of $T^3$.

     To study this, let us focus on the spectrum of particles, and
their 3-brane duals, in seven dimensions.  From the M-theory point of
view, they arise from membranes and 5-branes wrapping around the
2-cycles of K3.  In our discussion given in section \ref{sec:k3}, we
encountered two different types of 2-cycle in the construction that we
were using for K3, namely six 2-cycles corresponding to the usual
non-contractible 2-surfaces in $T^4$, and sixteen additional 2-cycles,
each associated with one of the sixteen Eguchi-Hanson instantons used
to smooth out the conical singularities on the identified 4-torus.
>From the field-theoretic point of view, the concept of ``wrapping'' a
$p$-brane soliton around a particular $m$-cycle in the internal space
essentially translates into the idea of constructing a $(p-m)$-brane
soliton supported by the lower-dimensional field arising in the
Kaluza-Klein expansion from the harmonic form associated to the given
$m$-cycle on the internal space.  When $m$ is less than the dimension
of the internal manifold, part of the transverse space of the
$p$-brane also becomes internal.  In a toroidal reduction, this notion
of vertical reduction can be made mathematically precise, by
``stacking'' $p$-branes along the reduction axes.  No such analogous
procedure has yet been implemented for K3 or Calabi-Yau
compactifications, where the internal coordinates correspond both to
the world-volume and the transverse space of the wrapped $p$-brane.
(Such an implementation would presumably first of all require that one
know the explicit metric on the K3 or Calabi-Yau internal space.)

   When the membrane wraps around any of the six 4-torus 2-cycles, it
gives a corresponding set of six 0-branes that are supported by the
six 2-form field strengths coming from the Kaluza-Klein reduction of
the $D=11$ 4-form on $T^4$.  These correspond, in the heterotic
picture, to the six 0-branes that describe the three winding and three
Kaluza-Klein modes on the compactifying 3-torus.

When the membrane wraps instead around any of the sixteen 2-cycles
associated with the Eguchi-Hanson manifolds, it should give rise to
sixteen 0-branes that are supported by the associated 2-form field
strengths which, in the heterotic picture, come from the $T^3$
dimensional reduction of the Yang-Mills fields of the $D=10$ heterotic
string.  However, a standard construction of a $p$-brane supported by
one of the Yang-Mills fields leads to a solution that is not
supersymmetric.\footnote{By the ``standard construction'' we mean that
one starts with a Lagrangian of the form $e^{-1}\, {\cal L} = R -\ft12
(\del\phi)^2 - e^{a\phi}\, F^2$, and makes a metric ansatz of the
usual form $ds^2 = e^{2A(r)}\, dx^\mu\, dx_\mu + e^{2B(r)}\, (dr^2 +
r^2\, d\Omega^2)$, with $F$ either of the form $F = Q\, \Omega$ (for a
magnetic solution) or of the dual form (for an electric solution).}
The reason why such a $p$-brane is non-supersymmetric is that the
supersymmetry transformation rule for the gauginos is of the form
$\delta\lambda^\I \sim G^\I_{\mu\nu}\, \Gamma^{\mu\nu}\, \ep$, and the
matrix $G^\I_{\mu\nu}\, \Gamma^{\mu\nu}$ is non-degenerate in the case
of the standard ansatz for one of the fields $G_\2^\I$.

    On the other hand, we know that the Cartan-subalgebra $U(1)^{16}$
subset of the Yang-Mills fields, together with the Kaluza-Klein and
winding vector fields, form an irreducible multiplet in the
fundamental representation of $O(10-D,26-D)$.  Thus naively one might
expect that starting from a supersymmetric $p$-brane supported either
by a Kaluza-Klein vector or by a winding vector, one could rotate to a
solution supported purely by the Yang-Mills fields.  However, it turns
out that this is not the case; the reason is that solutions supported
only by the Yang-Mills fields would lie on different $O(10-D,26-D)$
orbits from those supported by a single Kaluza-Klein or winding vector
field.  In this section we shall study the various possible types of
orbit that can arise, in order to see on which orbits supersymmetric
solutions supported purely by the Yang-Mills fields can lie.

\subsection{$D=9$ black holes and Yang-Mills wave excitations
\label{ssec:d9electric}}

     The key issues can be adequately illustrated by considering the
nine-dimensional theory arising from the dimensional reduction of
$N=1$ supergravity in $D=10$ coupled to a single $D=10$ Maxwell
multiplet.  This subset of the heterotic-theory fields has a global
$O(1,2)$ invariance in $D=9$.  The bosonic sector of the Lagrangian is
given by (\ref{d9lag}), with the index $I$ on the Yang-Mills fields
$B^I_\1$ and $B^\I_{\0}$ taken to have one value only.  We shall
denote the associated potentials by $B_\1$ and $B_\0$.  The Lagrangian
is
\bea
e^{-1}\, {\cal L}_9 &=& R -\ft12 (\del\phi)^2
-\ft12(\del\varphi)^2 - \ft12 e^{\sqrt2\varphi}\,
 (\del B_\0)^2 -\ft1{12} e^{-\sqrt{\ft87}\phi}\, (F_\3)^2
\nn\\
&& -\ft14 e^{-\sqrt{\ft27}\phi}\, \Big( e^{\sqrt2 \varphi}\,
(F_\2)^2 + e^{-\sqrt2 \varphi}\, (\cF_\2)^2 +
(G_\2)^2 \Big)\ ,\label{d9lag3}
\eea
 where the various field strengths, following (\ref{newfields}), are
given by
\bea
&&\cF_\2= d\cA_\1\ ,\quad G_\2=dB_\1 + B_\0\, d\cA_\1 \ ,\quad
  F_\2 = dA_\1 + B_\0\, dB_\1 + \ft12(B_\0)^2 \, d\cA_\1\ ,\nn\\
&&F_\3 = dA_\2 + \ft12 B_\1\, dB_\1 -\ft12 \cA_\1\, dA_\1 -\ft12
A_\1\, d\cA_\1\ .\label{d9fields}
\eea

    We can construct extremal 2-charge electric 0-brane or magnetic
5-brane solutions using the Kaluza-Klein and winding vectors $\cA_\1$
and $A_\1$.  In this subsection, we shall look at the
electrically-charged situation, and shall consider extremal black-hole
solutions.  The non-vanishing fields in this case are given by
\bea
&&ds_9^2 = (H_1\, H_2)^{-6/7}\, dx^\mu\, dx_\mu + (H_1\, H_2)^{1/7}\,
    (dr^2 + r^2\, d\Omega_7^2)\ ,\nn\\
&&e^{-\sqrt{14}\, \phi} = H_1\, H_2\ ,\qquad e^{\sqrt2\, \vp} =
\fft{H_1}{H_2} \ ,\label{2chargesole}\\
&& A_\1 = \pm H_1^{-1}\, dt\ ,\qquad \cA_\1 = \pm 
H_2^{-1}\, dt\ ,\qquad B_\1=0\ ,
\nn
\eea
 where the harmonic functions are given by\footnote{For simplicity, we
are considering single-centre isotropic solutions here.  The
discussion that follows can be immediately generalised to multi-centre
solutions, where $H_i = 1 +\sum_a \fft{q_i^a}{|\vec y-\vec y_a^i|}$,
with the potential $\omega$ written in terms of ${*dH_i}$ in the
standard way.}
\be
H_1 = 1 + \fft{q_1}{r^6}\ ,\qquad H_2 = 1 + \fft{q_2}{r^6}\ .\label{harme}
\ee
 The $\pm$ signs on the 1-form potentials in (\ref{2chargesole}) are
independent, and reflect the fact that the bosonic equations are
quadratic in field strengths, and hence solutions exist for any
choice of signs.  The electric charges $Q_1$ and $Q_2$ are given by
\be
Q_1 = \pm q_1\ ,\qquad Q_2 = \pm q_2\ .\label{q1q2e}
\ee
 The mass of the black hole is given by
\be
m = q_1 + q_2\ .\label{d9masse}
\ee

     Although the bosonic theory has four solutions, corresponding to
the four different sign choices in (\ref{2chargesole}), the
supersymmetry transformations depend linearly on the field strengths,
and, consequently, not all of the four sign choices for the charges
need yield supersymmetric solutions. In the case of maximal
supergravities, on the other hand, the fraction of preserved
supersymmetry {\it is} independent of the sign choices for similar
2-charge solutions. This can be seen from the fact that the 32
eigenvalues of the \bog matrix are \cite{lpsol} $\mu = m\pm Q_1 \pm
Q_2$, with each of the four sign combinations occurring with
multiplicity 8.  Thus in maximal supergravity, the 2-charge solutions
always preserve $\ft14$ of the supersymmetry, regardless of the choice
of signs.  On the other hand, in $N=1$ nine-dimensional supergravity
the 16 eigenvalues of the corresponding \bog matrix are a subset of
the 32 given above. Specifically, they are given by
\be
\mu = m \pm (Q_1 + Q_2)\ ,\label{d9bogo}
\ee
 with each sign choice occurring with multiplicity 8.
This means that in the $N=1$ theory, only two out of the four sign
choices in (\ref{q1q2e}) give solutions that preserve supersymmetry,
namely 
\be
(Q_1, Q_2) = (q_1,q_2)\ ,\qquad {\rm or} \qquad (Q_1, Q_2) = (-q_1,-q_2)
\ .\label{d9susy}
\ee
 In each of these cases, the 2-charge solution preserves $\ft12$ of
the $N=1$ supersymmetry.  For the remaining two sign choices in
(\ref{q1q2e}), all of the supersymmetry is broken.  From now on, we
shall consider just the supersymmetric solutions, and shall
concentrate on the first of the two cases listed in (\ref{d9susy}).
Note that when $Q_1=Q_2$ the two harmonic functions $H_1$ and $H_2$
become equal, and the solution reduces to a ``single-charge'' solution
supported by the single 2-form field strength of pure $N=1$
supergravity in $D=9$ \cite{lpss}.

    There are in total three electric 2-form charges $Q_1$, $Q_2$ and
$Q_3$ in the nine-dimensional theory that we are considering,
associated with the vectors $A_\1$, $\cA_\1$ and $B_\1$ respectively.
As we saw in section \ref{ssec:d=9het}, these vectors form a triplet
under the $O(1,2)$ global symmetry group.  The associated invariant
quadratic form constructed from the charges is
\be
{\cal I} = Q_3^2 + Q_{\sst Y}^2 - Q_{\sst X}^2 =
  Q_3^2 - 2Q_1\, Q_2\ ,\label{d9quad}
\ee
 where $Q_{\sst X} = \ft1{\sqrt2} (Q_1+Q_2)$, $Q_{\sst Y} =
\ft1{\sqrt2}( Q_1 -Q_2)$, as can be seen from the discussion in
section \ref{ssec:d=9het}.  The orbits on the three-dimensional charge
lattice that are filled out by acting on a given solution are thus
characterised by the quadratic invariant ${\cal I}$.  In particular,
there are three inequivalent types of orbit, corresponding to the cases
where ${\cal I}$ is positive, negative or zero.   

    To see the nature of the three types of orbit in detail, we first
note that a single-charge solution supported either by $Q_1$ or by
$Q_2$ belongs to a light-like orbit of $O(1,2)$, for which ${\cal I}$
vanishes.  On the other hand, if $Q_1$ and $Q_2$ are both
non-vanishing, and of the same sign, the orbit is time-like, with
${\cal I}<0$.  If instead these two charges have opposite signs, then
the orbit is space-like, with ${\cal I}>0$.

   Such solutions with light-like or time-like orbits have curvature
singularities which, for positive-mass solutions, lie on the horizon
at $r=0$, but they are free of singularities outside the horizon.  On
the other hand solutions lying on space-like orbits necessarily have
naked singularities that lie outside the horizon.  This is because the
charges are of opposite sign, and hence the supersymmetry requirement
(\ref{d9susy}) implies that one of the harmonic functions will have a
negative coefficient for its $r^{-6}$ term.  Although one might be
tempted to dismiss such solutions from consideration, their presence
is necessary in order to complete the charge lattice.  Indeed, a
single-charge solution supported exclusively by the Yang-Mills charge
$Q_3$ would lie on such a spacelike orbit.  Moreover, the fundamental
$D=9$ Yang-Mills field excitations of the theory corresponding to
spontaneously broken Yang-Mills generators must lie in short massive
supermultiplets that carry such charges.  Given the presumed
correspondence between such excitations and classical particle or wave
solutions, one would expect to find classical solutions with these
charges.  The fact that massive supermultiplets carrying Yang-Mills
charges must necessarily be short multiplets gives a hint that their
origin may be sought in massless wave-like super Yang-Mills solutions
in ten dimensions. Indeed, as we shall now show, they do originate
from such wave-like solutions, and moreover, this helps explain the
origin of the naked singularities.

    In fact, the naked singularity in the 2-charge solution
(\ref{2chargesole}) with $Q_1=q_1 > 0$ and $Q_2=q_2 <0$ is purely an
artefact of dimensional reduction.  If we oxidise the solution back to
$D=10$, the metric becomes
\be
ds_{10}^2 = H_1^{-3/4}\Big(- H_2^{-1}\, dt^2 + H_2\, (dz+
(H_2^{-1}-1)\, dt)^2 \Big) + H_1^{1/4}\, d\vec y\cdot d\vec y\ .
\ee
 This is the intersection of a string, associated with the harmonic
function $H_1$, and a gravitational pp-wave, associated with $H_2$.
The string has no associated naked singularity since $H_1$ has a
positive coefficient in its $r^{-6}$ term.  Although $r^{-6}$ in $H_2$
has a negative coefficient, this does not necessarily imply the
existence of a naked singularity in the associated wave solution.

    To study this issue, let us concentrate on a particular
solution lying on the space-like charge orbit which has a particularly
simple ten-dimensional interpretation.  If we act on the 2-charge solution
(\ref{2chargesole}) with the $O(1,2)$ transformation
\be
\Lambda = \pmatrix{1&0&0 \cr 0 & \cos\theta & \sin\theta 
        \cr 0 & -\sin\theta & \cos\theta}\ ,
\ee
 where $\tan\ft12\theta = (-q_1/q_2)^{1/2}$, we obtain a new solution
with the same metric as in (\ref{2chargesole}), but with the other
fields now given by
\bea
&&e^{-\sqrt{14}\, \phi} = H_1\, H_2 \ ,\qquad 
  e^{-\sqrt2\, \vp} = H_1\, H_2 \ ,\qquad B_0 = \fft1{\sqrt2} \, (H_2
- H_1) \, \sin\theta\ ,\nn\\
&& A_\1 =  ((H_1\, H_2)^{-1}-H_1^{-1} -H_2^{-1})\, dt\ ,
\qquad \cA_\1 = -(H_1\, H_2)^{-1}\, dt\ ,\nn\\
&&
B_\1 = \ft1{\sqrt2}\, \sin\theta\, (H_1^{-1} - H_2^{-1})\, dt\ .
\eea
 Oxidising this up to $D=10$, we obtain the solution
\bea
ds_{10}^2 &=& -(H_1\, H_2)^{-1}\, dt^2 + (H_1\,H_2) 
   (dz + (H_1\, H_2)^{-1}\, dt)^2 + d\vec y^2\ ,\nn\\
\phi_1 &=& 0\ ,\qquad B_\1 = \ft1{\sqrt2}\, \sin\theta\, (H_2-H_1)\,
dz\ ,\qquad A_\2 = {\rm pure\ gauge}\ .\label{d10wave}
\eea
 Note in particular that $F_\3=0$.  

    This solution describes a wave-like excitation of
the Yang-Mills field, propagating along the $z$ direction.  It can be
compared with the general class of wave-like solutions in the
heterotic string, described in \cite{guven}.  There, solutions were
sought of the form
\bea
&&ds_{10}^2 = -2 du\, (dv - W(u,\vec y)\, du)^2 + d\vec y^2\ ,\nn\\
&&B_1 = M(u,\vec y)\, du\ ,\qquad A_\2 = b(u,\vec y)\wedge du\ ,\qquad
\phi_1 = 0\ .\label{guvwave}
\eea
 For the class of solutions that are of interest to us, $b(u,\vec y)$ is
zero, and $M(u,\vec y)$ depends only on $\vec y$. From the results of
\cite{guven}, one has it that the equations of motion are satisfied
provided
\be
\del_i\, \del_i\, M = 0\ ,\qquad \del_i\, \del_i\, W = -\del_i\, M\,
\del_i\, M\ .
\ee
 Thus, $M$ is harmonic on the transverse space, and $W$ can be solved by
taking $W = \a\, M -\ft12 M^2$. If $M$ is the harmonic function $c\,
r^{-6}$, then the wave solution (\ref{guvwave}) is equivalent to
(\ref{d10wave}) under the following identifications:
\be
u=\fft{z}{\sqrt2}\ ,\qquad v= \fft{2t-z}{\sqrt2}\ ,\qquad 
c=\sqrt{-2q_1\, q_2}\ ,\qquad
\a= \fft{q_1+q_2}{\sqrt{-2q_1\, q_2}}\ .
\ee

     Although the $O(1,2)$ global symmetry can rotate the charges so
that the Kaluza-Klein electric charge $Q_2$ vanishes, one finds that the
Kaluza-Klein vector $\cA_\1$ can never become of pure gauge form upon
rotation from any non-trivial starting value. This implies that if a
generic $O(1,2)$ rotation of the solution (\ref{2chargesole}) is oxidised
to $D=10$, the metric will necessarily have a wave-like character; it is
given by
\be
ds_{10}^2 = (a^2\, H_1 + b^2\, H_2)^{-3/4}\, \Big( -2 dt + \fft{H_1\,
H_2}{a^2\, H_1 + b^2\, H_2}\, dz\Big)\, dz +(a^2\, H_1 + b^2\,
H_2)^{1/4}\ ,d\vec y^2\ ,
\ee
 where $a$ and $b$ are constants arising from the $O(1,2)$
rotation.  For generic $a$ and $b$, this solution describes the
intersection of a wave and a string.  The linear combination $a^2\,
H_1 + b^2\, H_2$ becomes the harmonic function associated with the
string.

     It is worth emphasising that the solutions with unavoidable naked
singularities in $D=9$ arise in the case where the quadratic invariant
(\ref{d9quad}) is positive, \ie where the charge orbit is space-like.
In \cite{sen0}, it was shown that for perturbative fundamental
single-string excitations, the above quadratic invariant ${\cal I}$ is
bounded above, lying in the interval\footnote{Note that our sign
convention is the opposite of the one used in \cite{sen0}.} $-\infty
<{\cal I} < 1$.  At the level of supergravity particle solutions, on
the other hand, there is no such upper bound.  This is because
supergravity solutions can describe not only single-string excitations
but also multiple-string excitations. The algebraic classification of
heterotic string states into single-string and multiple-string spectra
remains an interesting open problem.

\subsection{5-branes in $D=9$\label{ssec:d9magnetic}}

     Now we turn to the nine-dimensional 5-brane solution, which is given by
\bea
&&ds_9^2 = (H_1\, H_2)^{-1/7}\, dx^\mu\, dx_\mu + (H_1\, H_2)^{6/7}\,
    (dr^2 + r^2\, d\Omega_2^2)\ ,\nn\\
&&e^{\sqrt{14}\, \phi} = H_1\, H_2\ ,\qquad e^{\sqrt2\, \vp} =
\fft{H_2}{H_1} \ ,\label{2chargesol}\\
&& A_\1 = Q_1\, \omega\ ,\qquad \cA_\1 = Q_2\, \omega\ ,\qquad B_\1=0\ ,
\nn
\eea
 where the harmonic functions are given by
\be
H_1 = 1 + \fft{q_1}{r}\ ,\qquad H_2 = 1 + \fft{q_2}{r}\ ,\label{harm}
\ee
 and $d\omega=\Omega_\2$ is the volume form on the unit
2-sphere.  (Again, for simplicity we are considering single-centre
isotropic solutions here, which could easily be generalised to
multi-centre solutions.)  The mass per unit
4-volume of the 5-brane is given by
\be
m = q_1 + q_2\ .\label{d9mass}
\ee
 In terms of the two individual parameters $q_1$ and $q_2$, there exist
5-brane solutions where the magnetic charges under the Kaluza-Klein
and winding vectors are
\be
Q_1 = \pm q_1\ ,\qquad Q_2 = \pm q_2\ .\label{q1q2}
\ee
As in the case of electric black holes discussed earlier, the
supersymmetry requires that $(Q_1, Q_2) = (q_1, q_2)$ or
$(Q_1, Q_2) = (-q_1, -q_2)$.

    To construct a 5-brane solution supported purely by the Yang-Mills
2-form in $D=9$, let us begin with the solution (\ref{2chargesol}),
for which $Q_1=q_1$ and $Q_2=q_2$ are arbitrary, with $Q_1\, Q_2 <0$,
while $Q_3=0$. In order to map the charges to a configuration where
$Q_1=Q_2=0$, while $Q_3$ is non-zero, we must act with a matrix in an
$O(1,1)$ subgroup of $O(1,2)$ as follows:
\be
\pmatrix{\ft1{\sqrt2} (Q_1+Q_2) \cr \ft1{\sqrt2} (Q_1-Q_2) \cr  0} 
\longrightarrow  \pmatrix{ 0\cr 0\cr Q_3} =
          \pmatrix{\cosh t & \sinh t & 0 \cr
                         0 & 0 & -1 \cr
                         \sinh t & \cosh t & 0}\, 
  \pmatrix{\ft1{\sqrt2} (Q_1+Q_2) \cr \ft1{\sqrt2} (Q_1-Q_2) \cr  0} 
\ .
\ee
 The parameter $t$ is therefore given by
\be
e^{2t} = -\fft{Q_2}{Q_1}\ .
\ee
 Again, we see that $Q_1$ and $Q_2$ must have opposite signs in order
for the mapping to be possible.  Applying this transformation to the
fields $X$, $Y$ and $Z$ introduced in section \ref{ssec:d=9het}, one
can determine the transformed expressions for the dilaton $\vp$ and
the axion $B_\0$.  Similarly, by transforming the column vector
$(A_\1, \cA_\1, B_\1)$, one obtains the expressions for the
transformed vector potentials.  Upon doing this, we find that the
solution (\ref{2chargesol}) becomes
\bea
&&ds_9^2 = (H_1\, H_2)^{-1/7}\, dx^\mu\, dx_\mu + (H_1\, H_2)^{6/7}\,
    (dr^2 + r^2\, d\Omega_2^2)\ ,\nn\\
&&e^{\sqrt{14}\, \phi} = H_1\, H_2\ ,\qquad e^{\sqrt2\, \vp} =
\ft12 \sqrt{-\fft{Q_1\, H_2}{Q_2\, H_1}} + \ft12
\sqrt{-\fft{Q_2\, H_1}{Q_1\, H_2}}
 \ ,\label{2chargesol2}\\
&&\ft1{\sqrt2}\, B_\0 = \fft{Q_1\, H_2 + 
                             Q_2\, H_1}{Q_1\, H_2 - Q_2\, H_1}\ ,\nn\\
&& A_\1 = 0\ ,\qquad \cA_\1 = 0\ , \qquad 
B_\1 = Q_1\,  \sqrt{-\fft{2 Q_2}{Q_1}}\, \omega \ .\nn
\eea
 The harmonic functions retain their original form, and can in
general describe multi-centre solutions.  In the single-centre
isotropic case, they are given by (\ref{harm}).  In this special case,
the dilaton $\varphi$ and the axion $B_\0$ can be rewritten as
\be
e^{-2\sqrt2\varphi} = -\fft{4\, Q_1\, Q_2}{(Q_1-Q_2)^2}\, 
H_1\, H_2\ ,\qquad
\ft1{\sqrt2}\, B_\0 = \fft{Q_1\, H_2 + Q_2\, H_1}{Q_1-Q_2}\ .
\ee

     In line with our previous discussion we see indeed that this
solution, which carries only the charge $Q_3$ of the Yang-Mills field
strength $G_\2$, is not of the form of a ``standard'' single-charge
$p$-brane solution.  It is nevertheless, of course, supersymmetric,
since we have obtained it by performing an $O(1,2)$ rotation on a
standard supersymmetric solution.  Although it involves only a single
Yang-Mills charge $Q_3= q_1\sqrt{-2q_2/q_1}$, it has two independent
parameters $q_1$ and $q_2$ associated with the two harmonic functions
$H_1$ and $H_2$ given in (\ref{harm}).  However, the two parameters
$q_1$ and $q_2$ must both be non-zero and of opposite signs, implying
that there is again a naked singularity, at $r= |{\rm min}\,
(q_1,q_2)|$.  If $q_1=-q_2$, the 5-brane has zero mass.  The
connection between naked singularities and masslessness has been
extensively discussed in the context of BPS black holes of the $D=4$
toroidally-compactified heterotic string in Refs \cite{cvetic1,kl}. A
similar phenomenon also occurs in $p$-brane solitons in maximal
supergravity \cite{lpmult}, including massless dyonic strings in $D=6$
\cite{dlpdyon}.  The occurrence of a naked singularity means that the
naive relation between zero eigenvalues of the \bog matrix and
unbroken supersymmetries no longer holds, and in particular the
apparent enhancement of supersymmetry in the massless limit does not
in actuality occur \cite{dllp}.

     It is interesting to look at the form of the solution
(\ref{2chargesol2}) when oxidised back to $D=10$.  We find that it becomes
\bea
&&ds_{10}^2 = e^{\ft1{4\sqrt2}\vp}\, \Big( (H_1\, H_2)^{-1/8}\,
(dx^\mu\, dx_\mu + e^{-\sqrt2\vp}\, dz^2) + (H_1\, H_2)^{7/8}\, (dr^2
+ r^2\, d\Omega_2^2) \Big)\ ,\nn\\
&&e^{2\phi_1} = e^{\ft1{\sqrt2}\vp} (H_1\, H_2)^{-1/2} \ ,\qquad
A_\2 = Q_1\, \sqrt{-\fft{Q_2}{Q_1}}\, \Big( 
   \fft{Q_1\, H_2 + Q_2\, H_1}{Q_1\, H_2 -Q_2\, H_1}\Big)\,
   \omega\wedge dz\ ,\nn\\
&&B_\1 = Q_1\, \sqrt{-\fft{2Q_2}{Q_1}}\, \omega + \sqrt2\, 
 \Big( 
   \fft{Q_1\, H_2 + Q_2\, H_1}{Q_1\, H_2 -Q_2\, H_1}\Big)\, dz\ ,
\label{d10sol}
\eea
 where $e^{\sqrt2 \vp}$ is the function given in (\ref{2chargesol2}).
Note that in $D=10$ the solution also involves the NS-NS 3-form field
strength.  As usual, we give the metric in the Einstein frame here.
In terms of the string frame, this becomes
\be
ds_{\rm str}^2 =dx^\mu\, dx_\mu + e^{-\sqrt2\vp}\, dz^2 + 
(H_1\, H_2)\, (dr^2 + r^2\, d\Omega_2^2)\ .\label{d10stringmet}
\ee

     In the above discussion, we have considered the case where we
began with a 2-charge solution with $Q_1=q_1$ and $Q_2=q_2$ opposite
in sign.  By doing so, we ensured that the solution was
supersymmetric, but at the price of its having a naked singularity.
We could instead have started from a 2-charge solution with charges
$Q_1=q_1$ and $Q_2=-q_2$, again taken to be opposite in sign.  In this
case, the solution is non-supersymmetric, however, since now all the
\bog eigenvalues in (\ref{d9bogo}) are non-zero. However, if $Q_1$ and
$-Q_2$ are both positive then the solution will be free from naked
singularities.  This solution can also be rotated to one that carries
only the Yang-Mills charge $Q_3$.  The form of this solution is
identical to (\ref{2chargesol2}), except that the replacement of
$(Q_1,Q_2)=(q_1,q_2)$ by $(Q_1,Q_2)=(q_1,-q_2)$ implies that the
harmonic function $H_2$ is given by $1-Q_2/R$ rather than $1+Q_2/r$.
In particular, this means that when $Q_1=-Q_2$, and hence $H_1=H_2$,
the solution (\ref{2chargesol2}) is now a ``standard'' single-charge
5-brane solution, supported by the Yang-Mills field strength $G_\2$.
In line with our earlier discussion, this is indeed
non-supersymmetric.

This analysis of $p$-brane orbits supported by 2-form field strengths
in the toroidally-compactified heterotic string theory can be easily
generalised to lower dimensions.  In $D\ge 6$, the global symmetry
group is $O(10-D, 26-D)$, and the vector potentials form a
$(36-2D)$-dimensional representation.  The orbits are characterised by
an invariant quadratic form that generalises (\ref{d9quad}), and fall
into the three categories of time-like, space-like and light-like
orbits.  Any single-charge solution supported by a Kaluza-Klein vector
or a winding vector lies on a light-like orbit. A two-charge solution
involving both a Kaluza-Klein vector and a winding vector can lie
either on a time-like or a space-like orbit: When the two charges are
of the same sign, and hence the solution has no naked singularity, the
orbit is time-like.  It does not cover the points in the charge
lattice where the solution involves only charges that are carried by
the Yang-Mills 2-forms.  On the other hand, when the two charges are
of opposite signs, and so the solution suffers from a naked
singularity, the orbit is space-like and the solution can be rotated
to one supported purely by Yang-Mills field strengths.  All the above
solutions preserve half of the supersymmetry of the theory.  (There
are also time-like and space-like orbits for non-supersymmetric
two-charge solutions.  They can be obtained from the supersymmetric
2-charge solutions by reversing the relative sign of the two charges
but keeping the mass unchanged.  This phenomenon of supersymmetry
being broken as a consequence of a sign change of a charge, while
keeping the mass fixed, occurs also in maximal supergravities, but
only for solutions with more than 3 charges \cite{lpmult,ko,classp}.)

In $D=5$ there is an additional 2-form field strength that comes from
the dualisation of the NS-NS 3-form field.  This is a singlet under
the $O(5, 21)$ global symmetry, and there exists a 3-charge solution
involving this singlet, a Kaluza-Klein vector, and a winding vector.
In $D=4$, the global symmetry is enlarged to $O(6,24)\times SL(2,\R)$,
and the theory allows 4-charge solutions.  The solution space in this
case has been extensively studied \cite{klopp,dkmr,dr,cvetic2,cvetic3}.

In this section, we have been principally interested in supersymmetric
solutions supported purely by the Yang-Mills fields, in order to make
contact with M-theory compactified on K3. As we have discussed, such
solutions suffer from naked singularities. The origin of these naked
singularities in the M-theory picture is less clear. We have seen that
when an M-brane wraps around any of the sixteen Eguchi-Hanson 2-cycles
in K3, it gives rise to a $p$-brane which, in the heterotic picture,
is supported by a Yang-Mills field.  As we have seen in section
\ref{sec:k3}, these 2-cycles can shrink to zero size, in which case
the wrapped M-brane will become massless \cite{str}. Indeed, the
$p$-brane supported by the Yang-Mills field in the heterotic picture
can also become massless by adjusting the Yang-Mills moduli, \ie the
expectation values of the Cartan-subalgebra scalar fields.

\section*{Acknowledgments}

We are grateful to P.S. Aspinwall, G.W. Gibbons, C.J. Isham, 
R. Kallosh and A. Sen
for useful discussions.  H.L. and K.S.S. thank Texas A\&M University and
SISSA, and H.L. and C.N.P. thank CERN, for hospitality.

\section*{Appendices}
\addcontentsline{toc}{part}{Appendices}

\appendix

\section{Toroidal reduction of the heterotic string \label{app:hetred}}

   Our starting point is the low-energy effective Lagrangian for the
bosonic sector of the ten-dimensional heterotic string. In this paper, we have
considered the duality relations of the dimensionally reduced and ``fully
Higgsed'' heterotic string. That is, after dimensional reduction, we have
considered vacuum configurations in which the adjoint-representation
scalars arising in the dimensional reduction are assigned general expectation
values corresponding to vanishing potential energy. Such a general Higgs vacuum
causes a spontaneous breakdown of the $E_8\times E_8$ Yang-Mills gauge symmetry
to its maximal torus subgroup $U(1)^{16}$, whose algebra coincides 
with the Cartan
subalgebra of $E_8\times E_8$. It is thus the set of dimensional reductions of
this $U(1)^{16}$ Abelian Yang-Mills supergravity that principally concern us
here. For simplicity, we may begin by retreating to $D=10$ and
replacing the full
Yang-Mills-supergravity theory by its Abelian $U(1)^{16}$ maximal-torus
sub-theory. The bosonic sector of this $D-10$ Lagrangian is given by
\cite{BRWN}
\be
e^{-1}\, {\cal L}_{10} = R - \ft12 (\del\phi_1)^2 
 -\ft1{12} e^{\phi_1}\, (F_\3)^2  - \ft14 e^{\fft12 \phi_1}\, 
 \sum_I (G_\2^I)^2 \ ,\label{d10het}
\ee
 where $G_\2^I=dB_\1^I$ are the sixteen $U(1)$ gauge field strengths,
and $F_\3=dA_\2 +\ft12 B_\1^I\wedge dB_\1^I$.  The dilaton $\phi$ in
the standard convention is in fact  equal to $-\phi_1$. 

     We perform the toroidal reduction using the notation and conventions
of \cite{lpsol,cjlp1}.  The metric is reduced according to the ansatz
\be
ds_{10}^2 = e^{\vec s\cdot\vec\phi} \, ds_{\sst D}^2 +
\sum_{\a=2}^{11-\sst D} e^{2\vec\g_\a\cdot\vec\phi}\, (h^\a)^2\ ,
\label{dmetric}
\ee
 where $\vec \g_\a = \ft12\vec s - \ft12 \vec f_\a$ and
\bea
&&\vec s = (s_2,s_3,\ldots s_{11-\sst D}) \ ,\qquad
\vec f_\a = (\underbrace{0,0,\ldots,0}_{\a-2}, (10-\a)\, s_\a,
s_{\a+1}, s_{\a+2},\ldots, s_{11-\sst D})\ ,\nn\\
&&s_\a = \sqrt{\fft{2}{(10-\a)(9-\a)}}\ ,\qquad 2\le \a\le 11-D\ .
\eea
 The vectors $\vec s$ and $\vec f_\a$ satisfy
\be
\vec s\cdot\vec s = \fft{10-D}{4(D-2)}\ ,\qquad \vec s\cdot \vec f_\a
=  \fft2{D-2}\ ,\qquad
\vec f_\a\cdot\vec f_\b = 2 \, \delta_{\a\b} + \fft2{D-2}\ .\label{sfdot}
\ee
 Here, we are using $\vec\phi$ to denote the $(10-D)$-component vector
$\vec\phi = (\phi_2,\phi_3,\ldots, \phi_{11-\sst D})$.  Note that we
have reserved $\phi_1$ as the (negative of) the original dilaton in
$D=10$.  The quantities $h^\a$ are given by
\be
h^\a = dz^\a + \cA_\1^\a + \cA^\a_{\0\b}\, dz^\b = \td\g^\a{}_\b\, 
    (dz^\b + \hA_\1^\b)\ ,
\ee
 where $\td\g^\a{}_\b = \delta^\a_\b +\cA^\a_{\0\b}$, and
$\gamma^\a{}_\b$ is the inverse of $\td\g^\a{}_\b$ \cite{lpsol,cjlp1}.
We have also introduced the redefined Kaluza-Klein vectors $\hA_1^\a =
\g^\a{}_\b\, \cA_\1^\b$.  Note that $\g^a{}_\b$ and $\td\g^\a{}_\b$ are
non-zero only when $\a\le\b$. 

   Applying the above reduction ansatz, we obtain the $D$-dimensional
Lagrangian
\bea
e^{-1}\, {\cal L}_{\sst D} &=& R  -\ft12
(\del \vec\phi)^2 -\ft1{12} e^{\vec a_1\cdot\vec\phi}\, (F_\3)^2 -
\ft14 \sum_\a e^{\vec a_{1\a}\cdot\vec\phi}\,  (F_{\2\a})^2 \nn\\
&&
-\ft12 \sum_{\a<\b} e^{\vec a_{1\a\b}\cdot\vec\phi}\, 
(F_{\1\a\b})^2  -\ft14 \sum_I e^{\vec c\cdot\vec\phi}\, 
 (G_\2^I)^2 -\ft12 \sum_{\a, I} e^{\vec c_{\a}\cdot\vec \phi} \, 
 (G_{\1\a}^I)^2 \nn\\
&& -\ft14 \sum_{\a} e^{\vec b_\a\cdot\vec\phi}\, 
(\cF_\2^\a)^2  -\ft12 \sum_{\a<\b} e^{\vec b_{\a\b}\cdot\vec\phi}\, 
 (\cF^\a_{\1\b})^2 \ .\label{ddimhet}
\eea
 The indices $\a,\b\,\ldots$ run from 2 to $11-D$.
The various field strengths here are given by
\bea
F_\3 &=& dA_\2 +\ft12 B_\1^I\, dB_\1^I -(dA_{\1\a} +\ft12 B_{\0\a}^I\,
dB_\1^I +\ft12 B_\1^I\, dB_{\0\a}^I)\, \hA_\1^\a\nn\\
&& +\ft12(dA_{\0\a\b} -B^I_{\0[\a}\, dB^I_{\0\b]})\hA_\1^\a\,
\hA_\1^\b\ ,\nn\\
F_{\2\a}&=&\gamma^\b{}_\a\, \Big( dA_{\1\b} + \ft12 B^I_{\0\b}\, dB_\1^I 
 +\ft12 B_\1^I\, dB^I_{\0\b} +(dA_{\0\b\g} - B^I_{\0 [\b}\, dB^I_{\0
   \g]}) \, \hA_\1^\g\Big)\ ,\nn\\
F_{\1\a\b} &=& \g^\g{}{}_\a\, \g^\delta{}_\b\, (dA_{\0 \g\delta} -
B^\I_{\0 [\g}\, dB^\I_{\0 \delta]})\ ,   \label{fields}\\
G^I_\2 &=& dB_\1^I - dB^I_{\0\a}\, \hA_\1^\a\ ,\nn\\
G^I_{\1\a} &=& \g^\b{}_\a\, dB^\I_{\0\b}\ ,\qquad
\cF_\2^\a = \td\gamma^\a{}_\b\, d\hA_\1^\b\ ,\qquad
\cF^\a_{\1\b} = \gamma^\gamma{}_\b\, d\cA^\a_{\0\g}\ ,\nn
\eea

In the Lagrangian (\ref{ddimhet}), we have augmented the
$(10-D)$-component vector $\vec\phi$ introduced in (\ref{dmetric}) by
appending $\phi_1$ as its first component, so that now $\vec\phi =
(\phi_1,\phi_2,\ldots,\phi_{11-\sst D})$.  The dilaton vectors $\vec
a_1, \ldots$ characterise the couplings of the dilatonic scalars
$\vec\phi$ to the various field strengths; their details can be found
in \cite{lpsol,cjlp1}, and they are given by
\bea
&&\vec a_1 = (1,-2\, \vec s)\ ,\qquad 
    \vec a_{1\a} = (1,\vec f_\a-2\, \vec s)\ ,  
\qquad \vec a_{1\a\b} = (1, \vec f_\a + \vec f_\b - 2\, \vec s)\ ,\nn\\
&&\vec b_\a = (0,-\vec f_\a)\ ,\qquad 
 \vec b_{\a\b} = (0,-\vec f_\a + \vec f_\b)\ ,\nn\\
&&\vec c= (\ft12, -\vec s) = \ft12 \vec a_1\ ,\qquad
 \vec c_\a = (\ft12, \vec f_\a - \vec s) \ .\label{abcvecs}
\eea
 These dilaton vectors $\vec c_\a$ satisfy the orthogonality property
$\vec c_\a\cdot\vec c_\b = 2\, \delta_{\a\b}$. The dot product of
$\vec a_1$ or $\vec c$ with $\vec b_{\a\b}$ vanishes.  In fact $\vec
b_{\a\b}$ can also be written as $\vec b_{\a\b} = -\vec c_\a + \vec
c_\b$.  For the 3-form $F_\3$ and its dimensional reductions
$F_{\2\a}$ and $F_{\1\a\b}$, the dilaton vectors $\vec a_{1}$, $\vec
a_{1\a}$ and $\vec a_{1\a\b}$ are precisely those introduced in
\cite{lpsol,cjlp1} for the dimensional reduction of the 4-form field
$F_\4$ in $D=11$ supergravity.  (The 3-form $F_\3$ here coincides with
the NS-NS 3-form of type IIA supergravity, which arises as the
first-step reduction of $F_\4$.)  The dilaton vectors $\vec b_\a$ and
$\vec b_{\a\b}$, for the Kaluza-Klein vectors $\cA_\1^\a$ and the
Kaluza-Klein axions $\cA^\a_{\0\b}$ also coincide precisely with the
ones introduced in \cite{lpsol,cjlp1}.  The reason for these relations
to the dilaton vectors arising from the reduction of $D=11$
supergravity is that the pure supergravity sector of the heterotic
theory can be obtained as a truncation of type IIA supergravity where
the R-R fields are set to zero.  Finally, the dilaton vectors $\vec c$
and $\vec c_\a$ for the Yang-Mills fields $G_\2^I$ and their
dimensional reductions $G^I_{\1\a}$ come from reducing 2-form field
strengths in $D=10$ to 2-forms and 1-forms in $D$ dimensions.

    It is advantageous at this stage to perform some field
redefinitions on the 2-form and 1-form vector potentials, in order to
simplify the structure of the equations.  Thus we define
\bea
B_\1^\I &=& {B'}_\1^\I + B^\I_{\0\a}\, \hA_\1^\a\ ,\nn\\
A_{\1\a} &=& A'_{\1\a} - A_{\0\a\b}\, \hA_\1^\b + \ft12 B^\I_\a\,
{B'}_\1^\I \ ,\label{dredefs}\\
A_\2 &=& A_\2' + \ft12 A_{\0\a\b} \, \hA_\1^\a\, \hA_\1^\b + \ft12
B^\I_{\0\a}\, {B'}_\1^\I\, \hA_\1^\a + \ft12 A'_{\1\a}\, \hA_\1^\a\ .\nn
\eea
 In terms of these, the 3-form and 2-form field strengths become
\bea
F_\3 &=& dA_\2' + \ft12 {B'}_\1^\I\, d{B'}_\1^\I - \ft12 \hA_\1^\a\, 
dA'_{\1\a} - \ft12 A'_{\1\a}\, d\hA_\1^\a \ ,\nn\\
F_{\2\a} &=& \g^\b{}_\a\, \Big( dA'_{\1\b} - A_{\0\b\g}\, d\hA_\1^\g +
B^\I_{\0\b} \, d{B'}_\1^\I +\ft12 B^\I_{\0\b}\, B^\I_{\0\g}\,
d\hA_\1^\g \Big) \ ,\label{newfields}\\
G^\I_\2 &=& d{B'_\1}^\I + B^\I_{\0\a}\, d\hA_\1^\a\ ,\nn \\
\cF_\2^\a &=& \td\g^\a{}_\b\, d\hA_\1^\b\ .\nn
\nn
\eea
 The expressions for the 1-form field strengths $F_{\1\a\b}$,
$\cF^\a_{\1\b}$ and $G^\I_{\1\a}$ are unchanged, and still given by  
(\ref{fields}).

\section{Further coset geometry examples \label{app:morecos}}

     In section \ref{ssec:grassman} of this paper, we gave a
geometrical construction of the class of cosets (\ref{cosets}), which
includes those arising in the toroidal dimensional reduction of
ten-dimensional simple supergravity coupled to $N$ abelian gauge
fields. In this appendix, we extend this construction to several
further classes of cosets.  Since the principles of the construction
are closely parallel to those discussed in section
\ref{ssec:grassman}, we shall just present the essential features of
the examples.

   We begin with an additional class of cosets in which the matrix $W$
is real, and satisfies the pseudo-orthogonality condition $W^\T\,
\eta\, W = \eta$.  Then we can have an antisymmetric-matrix embedding:

\bigskip
\noindent\underline{$W^\T\, \eta\, W = \eta\ ,\qquad W^\T = -W$}
\bigskip

The numerator group here is generated by matrices $\Lambda$ satisfying
$\Lambda^\T\, \eta\, \Lambda=\eta$, whose action on $W$ is
$W\rightarrow \Lambda^\t\, W\, \Lambda$.  
It is easily seen that, without loss of generality, the fiducial
matrix $W_0$ can be taken to have the antisymmetric block diagonal
form
\be
W_0 = {\rm diag}\, (\sigma,\sigma,\ldots, \sigma)\ ,
\ee
 where $\sigma$ is the $2\times 2$ antisymmetric matrix
\be
\sigma= \pmatrix{0 & 1\cr -1 & 0}\ .\label{sigmamatrix}
\ee
 It is then evident that this matrix can only satisfy the given
conditions if $\eta$ has the form 
\be
\eta ={\rm diag}\, (\underbrace{-1,-1,\ldots, -1}_{2p}, 
                  \underbrace{1,1,\ldots, 1}_{2q} )\ .
\ee
 Thus we see that the numerator group is $O(2p,2q)$, while the stability
subgroup is given by $O(2p,2q)$ matrices $\Lambda$ that satisfy not
only $\Lambda^\T\, \eta\, \Lambda=\eta$ but also $\Lambda^\T\, W_0\,
\Lambda = W_0$.  It is not hard to see that with these conditions, the
matrices $\Lambda$ lie in $U(p,q)$.  Thus we obtain that the $O(2p,2q)$
orbits span the coset space
\be
\fft{O(2p,2q)}{U(p,q)}\ .
\ee

   For the next examples, we consider two classes of coset where $W$ is
real, but now leaves invariant the antisymmetric metric $\Omega$ defined
by
\be
\Omega = {\rm diag}\, (\sigma, \sigma, \ldots, \sigma)\ ,\label{omegadef}
\ee
 where $\sigma$ is given by (\ref{sigmamatrix}).  In these cases the
numerator group is generated by matrices $\Lambda$ that satisfy
$\Lambda^\T\, \Omega\, \Lambda = \Omega$, and whose action on $W$ is
$W\rightarrow \Lambda^\T\, W\, \Lambda$. We first consider a
symmetric-matrix embedding:

\bigskip
\noindent\underline{$W^\T\, \Omega\, W = \Omega\ ,\qquad W^\T=W$}
\bigskip

   In this case, the numerator group is defined by matrices $\Lambda$
satisfying $\Lambda^\T\, \Omega\, \Lambda=\Omega$, and thus it is
$Sp(2n,\R)$.  It is easy to see that the fiducial matrix $W_0$ can be
taken to be diagonal, and of the form
\be
W_0 = (\underbrace{-1,-1,\ldots, -1}_{2p}, 
       \underbrace{1,1,\ldots, 1}_{2q})\ .
\ee
 (The individual numbers of $-$ and $+$ signs must necessarily both be
even, in order that $W_0$ be able to satisfy the given conditions.)
The stability subgroup is given by matrices $\Lambda$ that satisfy
both $\Lambda^\T\, \Omega\, \Lambda = \Omega$ and $\Lambda^\T\, W_0\,
\Lambda= W_0$.  These are the same conditions as in the previous
example, and so again we find that the stability subgroup to be
$U(p,q)$.  Thus we obtain the coset
\be
\fft{Sp(2p+2q,\R)}{U(p,q)}\ .
\ee

     Next, consider the antisymmetric-matrix embedding:

\bigskip
\noindent\underline{$W^\T\, \Omega\, W = \Omega\ ,\qquad W^\T=-W$}
\bigskip

   The numerator group is the same as for the symmetric-matrix embedding
above. The fiducial matrix should be taken to be antisymmetric in this
case; we may pick
\be
W_0 ={\rm diag}\, (\underbrace{-\sigma, -\sigma, \ldots, -\sigma}_{p},
 \underbrace{\sigma, \sigma, \ldots, \sigma}_{q})\ .
\ee
 The stability subgroup is defined by matrices $\Lambda$ satisfying
both $\Lambda^\T\, \Omega\, \Lambda = \Omega$ and $\Lambda^\T\,W_0\,
\Lambda = W_0$, and it is thus clearly $SP(2p,\R)\times SP(2q,\R)$.
The coset in this case is therefore
\be
\fft{Sp(2p+2q,\R)}{Sp(2p,\R)\times Sp(2q,\R)}\ .
\ee

    Continuing on with metric-preserving numerator groups, we can now
consider cases where the numerator group is generated by complex
matrices $\Lambda$ that preserve the sesquilinear form
$\Lambda^\dagger\, \eta\, \Lambda=\eta$.  These act on complex
matrices $W$ according to $W\rightarrow \Lambda^\dagger\, W\, \Lambda$.  
Without loss of generality, $\eta$ can be taken to be symmetric. Thus we
may consider a Hermitean embedding:

\bigskip
\noindent\underline{$W^\dagger\, \eta\, W =\eta\ ,\qquad
  W^\dagger=W$:}
\bigskip

   Here, the metric $\eta$ has the form
\be
\eta= {\rm diag}\, (\underbrace{-1,-1,\ldots, -1}_p, 
     \underbrace{1,1,\ldots, 1}_q)\ .
\ee
 Note that there is no loss of generality in taking $W$ to be
Hermitean rather than anti-Hermitean, since the latter is related to
the former by a multiplication by $\im$.  The fiducial matrix can be
taken to be
\be
W_0 = {\rm diag}\, (\underbrace{-1,-1,.\ldots, -1}_m, 
        \underbrace{1,1,\ldots, 1}_{p-m}, 
       \underbrace{-1,-1,.\ldots, -1}_n, 
        \underbrace{1,1,\ldots, 1}_{q-m})\ .
\ee
 The numerator group is $U(p,q)$, and its stability subgroup is
$U(m,q-n)\times U(p-m,n)$.  Thus we have the coset
\be
\fft{U(p,q)}{U(m,q-n)\times U(p-m,n)}\ .
\ee

    As a final set of examples, we consider two cases where the
numerator group is not metric-preserving.  Specifically, we shall take
it to be $SL(n,\R)$, defined by $n\times n$ real matrices that satisfy
$\det \Lambda =1$. Their action on $W$ is $W\rightarrow \Lambda^\T\,
W\, \Lambda$. We can then take the coset matrices $W$ to satisfy
either $W^\T=W$ or $W^\T=-W$:

\bigskip
\noindent\underline{$W^\T= W$}
\bigskip

  The fiducial matrix $W_0$ can be taken to be
\be
W_0= {\rm diag}\, (\underbrace{-1,-1,\ldots, -1}_p, 
     \underbrace{1,1,\ldots, 1}_q)\ .
\ee
 It follows that the stability subgroup will be the subgroup
of $SL(p+q,\R)$ matrices that satisfy $\Lambda^\T\, W_0\, \Lambda =
W_0$; in other words it will be $SO(p,q)$.  Thus we have the coset
\be
\fft{SL(p+q,\R)}{SO(p,q)}\ .
\ee

\bigskip
\noindent\underline{$W^\T= -W$:}
\bigskip

 The fiducial matrix $W_0$ can be taken to be
\be
W_0= {\rm diag}\, (\sigma, \sigma,\ldots, \sigma)\ .
\ee
 The stability subgroup will therefore be generated by the subset
of $SL(2n,\R)$ matrices that satisfy $\Lambda^\T\, W_0\, \Lambda =
W_0$; in other words it will be $Sp(2n,\R)$.  Thus we have the coset
structure
\be
\fft{SL(2n,\R)}{Sp(2n,\R)}\ .
\ee

\end{document}